\documentclass[a4paper,nofootinbib,prd]{revtex4-1}
\usepackage{amsmath}
\usepackage{graphicx}
\usepackage{epsfig}
\usepackage{amsfonts}
\usepackage{amssymb}
\usepackage{color}
\usepackage{float}
\usepackage{enumitem}
\usepackage{dcolumn}
\usepackage{hyperref}
\usepackage{array}
\usepackage{accents}
\usepackage[utf8]{inputenc}
\usepackage{mathtools}
\usepackage{tikz}
\usepackage{multirow}
\hypersetup{
    colorlinks=true,
    linkcolor=blue,
    filecolor=magenta,
    citecolor=red
}

\newcommand{\udt}[3]{#1^{#2}_{\phantom{#2}#3}}
\newcommand{\udut}[4]{#1^{#2\phantom{#3}#4}_{\phantom{#2}#3\phantom{#4}}}

\newcommand{\dut}[3]{#1_{#2}^{\phantom{#2}#3}}
\newcommand{\dudt}[4]{#1_{#2\phantom{#3}#4}^{\phantom{#2}#3}}

\newcommand{\lc}[1]{\accentset{\circ}{#1}}

\begin{document}

\title{Gravitational Wave Propagation and Polarizations in the Teleparallel analog of Horndeski Gravity}

\author{Sebastian Bahamonde}
\email{sbahamonde@ut.ee}
\affiliation{Laboratory of Theoretical Physics, Institute of Physics, University of Tartu, W. Ostwaldi 1, 50411 Tartu, Estonia}

\author{Maria Caruana}
\email{maria.caruana.16@um.edu.mt}
\affiliation{Institute of Space Sciences and Astronomy, University of Malta, Msida, Malta}
\affiliation{Department of Physics, University of Malta, Malta}

\author{Konstantinos F.	Dialektopoulos}
\email{kdialekt@gmail.com}
\affiliation{Center for Gravitation and Cosmology, College of Physical Science and Technology, Yangzhou University, Yangzhou 225009, China}

\author{Viktor Gakis}
\email{vgakis@central.ntua.gr}
\affiliation{Institute of Space Sciences and Astronomy, University of Malta, Msida, Malta}
\affiliation{Department of Physics, National Technical University of Athens, Zografou Campus GR 157 73, Athens, Greece}

\author{Manuel Hohmann}
\email{manuel.hohmann@ut.ee}
\affiliation{Laboratory of Theoretical Physics, Institute of Physics, University of Tartu, W. Ostwaldi 1, 50411 Tartu, Estonia}

\author{Jackson Levi Said}
\email{jackson.said@um.edu.mt}
\affiliation{Institute of Space Sciences and Astronomy, University of Malta, Msida, Malta}
\affiliation{Department of Physics, University of Malta, Msida, Malta}

\author{Emmanuel N. Saridakis}
\email{msaridak@phys.uoa.gr}
\affiliation{National Observatory of Athens, Lofos Nymfon, 11852 Athens, Greece}
\affiliation{Department of Astronomy, School of Physical Sciences, \\
University of Science and Technology of China, Hefei, Anhui 230026, China}
\affiliation{CAS Key Laboratory for Research in Galaxies and Cosmology, \\
University of Science and Technology of China, Hefei, Anhui 230026, China}

\author{Joseph Sultana}
\email{joseph.sultana@um.edu.mt}
\affiliation{Department of Mathematics, University of Malta, Msida, Malta}

\begin{abstract}
Gravitational waves (GWs) have opened a new window on fundamental physics in a number of important ways. The next generation of GW detectors may reveal more information about the polarization structure of GWs. Additionally, there is growing interest in theories of gravity beyond GR. One such theory which remains viable within the context of recent measurements of the speed of propagation of GWs is the teleparallel analogue of Horndeski gravity. In this work, we explore the polarization structure of this newly proposed formulation of Horndeski theory. In curvature-based gravity, Horndeski theory is almost synonymous with extensions to GR since it spans a large portion of these possible extensions. We perform this calculation by taking perturbations about a Minkowski background and consider which mode propagates. The result is that the polarization structure depends on the choice of model parameters in the teleparallel Horndeski Lagrangian with a maximum of seven propagating degrees of freedom. While the curvature-based Horndeski results follows as a particular limit within this setup, we find a much richer structure of both massive and massless cases which produce scalar--vector--tensor propagating degrees of freedom. We also find that the GW polarization that emerges from the teleparallel analogue of Horndeski gravity results in analogous massive and massless modes which take on at most four polarizations in the massless sector and two scalar ones in the massive sector. In none of the cases do we find vector polarizations.
\end{abstract}

\maketitle

\section{Introduction}
The observation of gravitational waves (GWs) first by the LIGO collaboration \cite{Abbott:2016blz} and now  also by the Virgo collaboration \cite{Abbott:2017oio} has made it possible to test general relativity (GR) in the strong field regime through template matching \cite{Carson:2020rea,Ghosh:2017gfp}. As with GR, theories beyond GR require a lot of theoretical and numerical work before competing templates can be produced for even the simplest scenarios such as binary black hole coalescence events. Due to these limitations, it may mean that GW polarization may be the most promising way forward in terms of testing gravitational theories beyond GR. Moreover, this provides a model-independent way to find the propagation of possible extra degrees of freedom (DoF) \cite{Ezquiaga:2018btd,Mukhanov:991646,Barack:2018yly}. \medskip

GWs propagate with only two tensor polarizations in GR \cite{Will:2005va} whereas a much richer polarization structure exists in many other theories of gravity. In general, theories of gravity that depend on Riemann manifolds can host a maximum of six polarizations which are related to the metric tensor form, namely two tensor polarization modes (plus and cross), two vector modes (vector $x$ and vector $y$), and two scalar modes (breathing and longitudinal) \cite{Will:2005va}. For instance, $f(\lc{R})$ gravity (over-circles refers to the quantities calculated with the regular Levi-Civita connection) propagates one extra scalar mode in certain scenarios in addition to the regular tensor polarizations \cite{DeFelice:2010aj,Liang:2017ahj,Gong:2017bru}, while Brans-Dicke theory just has an extra scalar breathing mode \cite{Eardley:1974nw}, whereas in bimetric gravity there is the possibility of having up to the maximum full six polarization modes \cite{Visser:1997hd}. On the other hand, theories exist which have identical GW polarization characteristics to GR such as $f(T)$ \cite{Bamba:2013ooa,Farrugia:2018gyz} and $f(Q)$ \cite{Soudi:2018dhv} gravity (where $T$ and $Q$ are  measures of torsion and nonmetricity, respectively) which are both non-Riemannian theories of gravity. \medskip

Preliminary work on the search for extra polarization modes has already started in earnest through the LIGO-Virgo collaboration since three detectors offer the possibility of cross-matching arrival time observations \cite{Abbott:2018utx}. This has been done in the context of black hole binary mergers in GW170814 \cite{Abbott:2017oio} and binary neutron star mergers in GW170817 \cite{TheLIGOScientific:2017qsa}. However, due to the theoretical limitations of template matching, the waveforms are assumed to be identical to GR and the vector and scalar modes searches have a strong element of model-dependence which constrains the applicability of present studies of this nature. These issues have been tackled in other studies \cite{Takeda:2018uai} but the matter may ultimately be resolved when more novel detectors come online. \medskip

GW polarizations offer the possibility of a clear deviation from GR that may be detectable by the next generation of detectors such as LISA \cite{Belgacem:2019pkk} and the Einstein Telescope \cite{Maggiore:2019uih}, among many others \cite{Miller:2019vfc}. In this work, we explore the GW polarization structure of the recently proposed teleparallel analogue of Horndeski gravity \cite{Bahamonde:2019shr}. Standard Horndeski gravity \cite{Horndeski:1974wa} is the most general second-order theory of gravity (in terms of metric derivatives) that contains only one scalar field, in the context of the Levi-Civita connection. However, other connections can also be used to describe different formulations of gravity. The Levi-Civita connection is characterized by curvature-based constructions of gravity. More broadly, a trinity of possible deformations to Riemann manifolds exists where not only curvature but also torsion and nonmetricity contribute \cite{BeltranJimenez:2019tjy}. In addition to the curvature-based Levi-Civita connection, Riemann manifolds are also compatible with a teleparallel connection, which is torsion-based \cite{Aldrovandi:2013wha}, and a disformation connection, which is based on nonvanishing nonmetricity \cite{BeltranJimenez:2017tkd}. \medskip

The recent measurements of the speed of propagation of GWs has constrained their speed to the speed of light to within deviations of at most one part in $10^{15}$ which has rendered severe limitations on the rich structure of standard Horndeski gravity \cite{Ezquiaga:2017ekz}. This has consequently drastically limited the number of viable paths to constructing physically realistic models of cosmology within standard Horndeski gravity. In Ref.~\cite{Bahamonde:2019shr}, Bahamonde-Dialektopoulos-Levi Said (BDLS) propose a teleparallel analogue of Horndeski gravity. Due to the naturally lower-order nature of teleparallel, or torsional-based gravity, the framework that ensues in BDLS theory is much richer than in standard gravity (i.e curvature-based gravity theories). One direct result of this feature is that the GW propagation speed in BDLS theory contains more terms related to the gravitational Lagrangian, as shown in Ref.~\cite{Bahamonde:2019ipm}. Crucially, this means that previously disqualified models of Horndeski gravity can be revived in a teleparallel gravity form through carefully chosen additional Lagrangian contributions. \medskip

The gravitational wave propagation equation gives the speed of propagation of GWs together with amplitude modulations which is addressed in Ref.~\cite{Ezquiaga:2018btd}. The next question one may ask is whether BDLS theory propagates further polarizations in addition to the standard Horndeski picture. In Ref.~\cite{Hou:2017bqj}, the polarization modes for different scenarios is explored in the context of standard Horndeski gravity. In this work, our central focus is targeted at the determination of the polarization modes in BDLS theory. In Sec.~\ref{sec:intro_BDLS} we briefly review the teleparallel gravity basis of BDLS theory and introduce the exact formulation of the teleparallel analogue of Horndeski gravity while in Sec.~\ref{sec:perturbations} the perturbation regime for this form of gravity s prescribed. In Sec.\ref{sec:dof_analysis} the perturbations are applied to BDLS theory and the propagating degrees of freedom (PDoF) are explored for various scenarios of the theory with a focus on the contributions from the decomposition in a scalar--vector--tensor (SVT) regime. Finally, the GW polarization classification is presented in Sec.~\ref{sec:gw_polarizations} which is followed by a discussion of the main results in Sec.~\ref{sec:conclusion}. Throughout the work, geometric units and the $(+1,-1,-1,-1)$ signature are used.

\section{The Teleparallel Analog of Horndeski Theory}\label{sec:intro_BDLS}
The curvature associated with GR is built on the Levi-Civita connection $\udt{\lc{\Gamma}}{\sigma}{\mu\nu}$ which is torsionless and satisfies the metric compatibility condition. Teleparallel gravity (TG) is based on torsion through the teleparallel connection, $\Gamma^{\sigma}{}_{\mu\nu}$, which continues to satisfy metricity but that is curvatureless. As is well known, in curvature based theories of gravity, the gravitational field is measured through the Riemann tensor and its contractions, which have led to a large number of extended theories of gravity beyond GR \cite{Clifton:2011jh,Capozziello:2011et}. This is a fundamental measure of curvature \cite{misner1973gravitation}. However, when the Levi-Civita connection is replaced with the teleparallel connection, the Riemann curvature tensor identically vanishes which points to the necessity of new measures of geometric deformation within the TG context. \medskip

The metric tensor, $g_{\mu\nu}$, is the fundamental object of GR and is defined as an inner product. In TG the metric tensor emerges from tetrad fields, $\udt{e}{A}{\mu}$, which act as soldering agents between the general manifold (Greek indices) and the local Minkowski space (capital Latin indices). This means that they can be used to raise Minkowski space indices to the general manifold and vice versa through \cite{RevModPhys.48.393}
\begin{align}
    g_{\mu\nu} &= \udt{e}{A}{\mu}\udt{e}{B}{\nu} \eta_{AB}\,,\label{eq:metr_trans}\\
    \eta_{AB} &= \dut{E}{A}{\mu}\dut{E}{B}{\nu} g_{\mu\nu}\,,
	\end{align} which also define the inverse tetrads, $\dut{E}{A}{\mu}$, which satisfy the dual basis conditions
\begin{align}
    \udt{e}{A}{\mu}\dut{E}{B}{\mu} &= \delta_B^A\,,\\
    \udt{e}{A}{\mu}\dut{E}{A}{\nu} &= \delta_{\mu}^{\nu}\,,
\end{align}
where $\eta_{AB}$ is the Minkowski metric. Since tetrads contain 6 extra DoFs compared to the metric, different tetrads can reproduce the same metric. This is manifested due to the local Lorentz transformations (LLTs), $\udt{\Lambda}{A}{B}$, on the local space. \medskip

In this context, TG is routed on the replacement of the Levi-Civita connection with the teleparallel connection which implies a corresponding exchange of the metric tensor with the tetrad as the fundamental dynamical object of the theory. Note that also GR can be formulated using tetrads, even though this is less common; in TG, however, tetrads are more commonly used than the alternative, but equivalent formulation in terms of a metric and an independent teleparallel connection~\cite{Hohmann:2021fpr}. The latter is the linear affine connection that satisfies both the curvatureless and metricity conditions \cite{Weitzenbock1923}, and denoted by the teleparallel connection. In the tetrad-based formulation, this connection can be expressed as \cite{Cai:2015emx,Krssak:2018ywd}
\begin{equation}
\Gamma^{\lambda}{}_{\nu\mu}=\dut{E}{A}{\lambda}\partial_{\mu}\udt{e}{A}{\nu}+\dut{E}{A}{\lambda}\udt{\omega}{A}{B\mu}\udt{e}{B}{\nu}\,,
\end{equation}
where $\udt{\omega}{A}{B\mu}$ represents the components of a flat spin connection, which must satisfy
\begin{equation}
    \partial_{[\mu}\udt{\omega}{A}{|B|\nu]} + \udt{\omega}{A}{C[\mu}\udt{\omega}{C}{|B|\nu]} \equiv 0\,.
\end{equation}
Given that this is a flat connection, the DoFs associated with the LLTs play an active role in the field equations of the theory such that the spin connection will counter balance these inertial effects so that the theory retains covariance. In any of these setups there will exist a frame in which the spin connection naturally vanishes, which is called the Weitzenb\"{o}ck gauge \cite{Krssak:2018ywd}. \medskip

TG is based on the torsion associated with the teleparallel connection. In this context, a torsion tensor must be defined to replace the the role that the Riemann tensor plays\footnote{It is important to emphasize that, in general, the Riemann tensor as calculated with Levi-Civita connection does not vanish $\udt{\lc{R}}{\alpha}{\beta\mu\nu} \neq 0$.}, whose components identically vanish. By definition, the torsion tensor is defined as the antisymmetric operator acted on the connection \cite{Aldrovandi:2013wha,ortin2004gravity}
\begin{equation}
    \udt{T}{A}{\mu\nu} := 2\udt{\Gamma}{A}{[\nu\mu]}\,,
\end{equation}
which acts as a measure of the field strength of gravitation in TG, and where the square brackets denote antisymmetric operator. The torsion tensor transforms covariantly under LLTs and is diffeomorphism invariant \cite{Krssak:2015oua}. An interesting feature of the torsion tensor is that it can be decomposed into three irreducible pieces, namely axial, vector and purely tensorial parts which are given by \cite{PhysRevD.19.3524,Bahamonde:2017wwk}
\begin{align}
a_{\mu} & :=\frac{1}{6}\epsilon_{\mu\nu\lambda\rho}T^{\nu\lambda\rho}\,,\\[4pt]
v_{\mu} & :=\udt{T}{\lambda}{\lambda\mu}\,,\\[4pt]
t_{\lambda\mu\nu} & :=\frac{1}{2}\left(T_{\lambda\mu\nu}+T_{\mu\lambda\nu}\right)+\frac{1}{6}\left(g_{\nu\lambda}v_{\mu}+g_{\nu\mu}v_{\lambda}\right)-\frac{1}{3}g_{\lambda\mu}v_{\nu}\,,
\end{align}
where $\epsilon_{\mu\nu\lambda\rho}$ is the totally antisymmetric Levi-Civita tensor in four dimensions. The irreducible parts of the torsion tensor all vanish when contracted with each other due to the symmetries of the torsion tensor itself. The decomposition of the torsion tensor can be used to form the three scalar invariants
\begin{align}
T_{\text{ax}} & := a_{\mu}a^{\mu} = -\frac{1}{18}\left(T_{\lambda\mu\nu}T^{\lambda\mu\nu}-2T_{\lambda\mu\nu}T^{\mu\lambda\nu}\right)\,,\\[4pt]
T_{\text{vec}} & :=v_{\mu}v^{\mu}=\udt{T}{\lambda}{\lambda\mu}\dut{T}{\rho}{\rho\mu}\,,\\[4pt]
T{_{\text{ten}}} & :=t_{\lambda\mu\nu}t^{\lambda\mu\nu}=\frac{1}{2}\left(T_{\lambda\mu\nu}T^{\lambda\mu\nu}+T_{\lambda\mu\nu}T^{\mu\lambda\nu}\right)-\frac{1}{2}\udt{T}{\lambda}{\lambda\mu}\dut{T}{\rho}{\rho\mu}\,,
\end{align}
that can be summed to form the torsion scalar, and which are parity preserving \cite{Bahamonde:2015zma}. These three scalars can be combined to form the most general purely gravitational Lagrangian containing only parity preserving scalars, $f(T_{\text{ax}},T_{\text{vec}},T{_\text{ten}})$, that is quadratic in contractions of the torsion tensor \cite{Bahamonde:2017wwk}, and where the resulting field equations are second order in derivatives of the tetrads. The torsion scalar can then be written as
\begin{equation}
T:=\frac{3}{2}T_{\text{ax}}+\frac{2}{3}T_{\text{ten}}-\frac{2}{3}T{_{\text{vec}}}=\frac{1}{2}\left(E_{A}{}^{\lambda}g^{\rho\mu}E_{B}{}^{\nu}+2E_{B}{}^{\rho}g^{\lambda\mu}E_{A}{}^{\nu}+\frac{1}{2}\eta_{AB}g^{\mu\rho}g^{\nu\lambda}\right)T^{A}{}_{\mu\nu}T^{B}{}_{\rho\lambda}\,,
\end{equation}
which is equivalent to the Ricci scalar, $\lc{R}$ (calculated using the Levi-Civita connection), less a total divergence term represented by \cite{Bahamonde:2015zma}
\begin{equation}
R=\lc{R}+T-\frac{2}{e}\partial_{\mu}\left(e\udut{T}{\lambda}{\lambda}{\mu}\right)=0\,,
\end{equation}
where $R$ is the Ricci scalar as calculated using the teleparallel connection, which organically vanishes due to the symmetry properties of the flat teleparallel connection and $e=\text{det}\left(\udt{e}{A}{\mu}\right)=\sqrt{-g}$ is the tetrad determinant. This means that we can define a boundary term, $B$, such that
\begin{equation}
\lc{R}=-T+\frac{2}{e}\partial_{\mu}\left(e\udut{T}{\lambda}{\lambda}{\mu}\right):=-T+B\,.
\end{equation}
Now, given the boundary term nature of $B$, the action produced by a linear contribution of $T$ alone is guaranteed to produce a teleparallel equivalent of general relativity (TEGR) \cite{Hehl:1994ue,Aldrovandi:2013wha}. An interesting feature of TEGR is that it can be formulated as native gauge theory of translations \cite{Aldrovandi:2013wha,Hayashi:1967se}. \medskip

In GR, the equivalence principle offers a procedure by which to raise local Lorentz frames to general ones by the exchange of the Minkowski metric with it general metric tensor while also raising partial derivatives to covariant ones (with respect to the Levi-Civita connection) \cite{misner1973gravitation,nakahara2003geometry}. TG is distinct in that this coupling prescription is largely preserved for additional fields while the gravitational connection is changed to the teleparallel connection \cite{Aldrovandi:2013wha}. Thus, the coupling prescription for this context is described by the raising of Minkowski tetrads to their arbitrary analogs, while the tangent space partial derivatives are raised to the regular Levi-Civita connection covariant derivatives, namely
\begin{equation}
    \partial_{\mu} \rightarrow \mathring{\nabla}_{\mu}\,,
\end{equation}
which emphasises the close relationship both theories have with each other \cite{BeltranJimenez:2019tjy}. \medskip

Now that both the purely gravitational and scalar field sectors have been adequately developed, we can consider the recently proposed teleparallel analog of Horndeski gravity \cite{Bahamonde:2019shr,Bahamonde:2019ipm,Bahamonde:2020cfv}. This new formulation of Horndeski gravity was built in TG using just three limiting conditions, namely that (i) the field equations must be at most second order in their derivatives of the tetrads; (ii) the scalar invariants will not be parity violating; and (iii) the number of contractions with the torsion tensor is limited to being at most quadratic. These conditions are necessary because TG is organically lower order in derivatives of the tetrad (since the torsion tensor depends only on first derivatives of the tetrad), and thus can produce a potentially infinite number of second order theories of gravity. This automatically implies a much weaker form of the Lovelock theorem \cite{Lovelock:1971yv,Gonzalez:2015sha,Gonzalez:2019tky}. Condition (iii) limits the number of such terms to a finite number, where such higher order contractions may play a role in other phenomenology. \medskip

The TG Horndeski conditions have been shown to lead directly to a finite number of contributing scalar invariants that explore possible nonminimal coupling terms between the scalar field and the torsion tensor, which for the linear case gives \cite{Bahamonde:2019shr}
\begin{equation}
    I_2 = v^{\mu} \phi_{;\mu}\,,
\end{equation}
while for the quadratic scenario gives
\begin{align}
J_{1} & =a^{\mu}a^{\nu}\phi_{;\mu}\phi_{;\nu}\,,\\[4pt]
J_{3} & =v_{\sigma}t^{\sigma\mu\nu}\phi_{;\mu}\phi_{;\nu}\,,\\[4pt]
J_{5} & =t^{\sigma\mu\nu}\dudt{t}{\sigma}{\alpha}{\nu}\phi_{;\mu}\phi_{;\alpha}\,,\\[4pt]
J_{6} & =t^{\sigma\mu\nu}\dut{t}{\sigma}{\alpha\beta}\phi_{;\mu}\phi_{;\nu}\phi_{;\alpha}\phi_{;\beta}\,,\\[4pt]
J_{8} & =t^{\sigma\mu\nu}\dut{t}{\sigma\mu}{\alpha}\phi_{;\nu}\phi_{;\alpha}\,,\\[4pt]
J_{10} & =\udt{\epsilon}{\mu}{\nu\sigma\rho}a^{\nu}t^{\alpha\rho\sigma}\phi_{;\mu}\phi_{;\alpha}\,,
\end{align}
where semicolons represent covariant derivatives with respect to the Levi-Civita connection. \medskip

The new scalar invariants can be arbitrarily combined to produce a new Lagrangian contribution to the regular Horndeski terms. The standard gravity Horndeski terms naturally emerge since TG is lower order, meaning that we at least find the regular contributions. The additional Lagrangian term appears as
\begin{equation}
    \mathcal{L}_{\text{Tele}}:= G_{\text{Tele}}\left(\phi,X,T,T_{\text{ax}},T_{\text{vec}},I_2,J_1,J_3,J_5,J_6,J_8,J_{10}\right)\,,
\end{equation}
where the kinetic term is defined as $X:=-\frac{1}{2}\partial^{\mu}\phi\partial_{\mu}\phi$. An aesthetic difference occurs for the standard terms in that they are now described by the tetrad rather than the metric but their values remain identical. Hence, the TG analog of Horndeski gravity is described by \cite{Bahamonde:2019shr,Horndeski:1974wa,Saridakis:2021lqd}
\begin{equation}\label{action}
    \mathcal{S}_{\text{BDLS}} = \frac{1}{2\kappa^2}\int d^4 x\, e\mathcal{L}_{\text{Tele}} + \frac{1}{2\kappa^2}\sum_{i=2}^{5} \int d^4 x\, e\mathcal{L}_i+ \int d^4x \, e\mathcal{L}_{\rm m}\,,
\end{equation}
where
\begin{align}
\mathcal{L}_{2} & :=G_{2}(\phi,X)\,,\label{eq:LagrHorn1}\\[4pt]
\mathcal{L}_{3} & :=G_{3}(\phi,X)\mathring{\Box}\phi\,,\\[4pt]
\mathcal{L}_{4} & :=G_{4}(\phi,X)\left(-T+B\right)+G_{4,X}(\phi,X)\left[\left(\mathring{\Box}\phi\right)^{2}-\phi_{;\mu\nu}\phi^{;\mu\nu}\right]\,,\\[4pt]
\mathcal{L}_{5} & :=G_{5}(\phi,X)\mathring{G}_{\mu\nu}\phi^{;\mu\nu}-\frac{1}{6}G_{5,X}(\phi,X)\left[\left(\mathring{\Box}\phi\right)^{3}+2\dut{\phi}{;\mu}{\nu}\dut{\phi}{;\nu}{\alpha}\dut{\phi}{;\alpha}{\mu}-3\phi_{;\mu\nu}\phi^{;\mu\nu}\,\mathring{\Box}\phi\right]\,,\label{eq:LagrHorn5}
\end{align}
where $\mathcal{L}_{\rm m}$ is the matter Lagrangian in the Jordan conformal frame, $\kappa^2:=8\pi G$, $\lc{G}_{\mu\nu}$ is the standard Einstein tensor, and comma represents regular partial derivatives. Standard Horndeski theory is clearly recovered when $G_{\text{Tele}} = 0$ is selected. Due to the invariance under LLTs and general covariance that stems from TG, the TG analog of Horndeski gravity organically adheres to these critical properties.

\section{Perturbations in Teleparallel Gravity}\label{sec:perturbations}

In this section we introduce our prescription for taking perturbations about a Minkowski background in the context of tetrads together with a scalar field. To do this, we expand the tetrad about the Minkowski background through
\begin{align}
    \udt{\tilde{e}}{A}{\mu} & =\delta_{\mu}^{A}+\epsilon\,\delta\udt{e}{A}{\mu}=\delta_{\mu}^{A}+\epsilon\,\delta_{B}^{\nu}\eta^{AB}\tau_{\mu\nu}\,,\label{eq:tetrad_pert}\\[4pt]
    \phi & =\phi_{0}+\epsilon\,\delta\phi\,,\label{eq:phi_pert}
\end{align}
where
\begin{align}
    \tau_{\mu\nu} & :=\eta{}_{AB}e^{B}{}_{\mu}\delta e{}^{A}{}_{\nu}\,,\label{eq:taumunu}
\end{align}
and where $\epsilon$ represents the order of the perturbation, $e^{B}{}_{\mu}$ represents the background value of the tetrad which is just $\delta^{B}{}_{\mu}$ for a Minkowski background, and $\phi_{0}$ is a constant (in general, the background value of $\phi_0$ can be taken to be time dependent as $\phi=\phi(t)$). Thus, the metric perturbation becomes \eqref{eq:metr_trans}
\begin{equation}\label{eq:metric_pert}
    \tilde{g}_{\mu\nu}=\eta_{\mu\nu}+\epsilon\,h_{\mu\nu}+\frac{1}{2}\epsilon^{2}\,\delta^{2}g_{\mu\nu}=\eta_{\mu\nu}+2\epsilon\,\tau_{(\mu\nu)}+\epsilon^{2}\,\tau_{\alpha\mu}\udt{\tau}{\alpha}{\nu}\,.
\end{equation}

To obtain the BDLS perturbation equations, we consider the perturbations of the action in Eq.~\eqref{action} which then results in the linearized equations of motion by taking an appropriate variation (see Appendix~\ref{Expansion_Multivariable_Function}). In the case of BDLS theory, Taylor expansion is performed about the point $\phi=\phi_{0}$ while the other arguments do not contribute at the background (since the background is the Minkowski space). Hence, we can write
\begin{align}
 & G_{\text{Tele}}\left(\phi,X,T,T_{\text{ax}},T_{\text{vec}},I_{2},J_{1},J_{3},J_{5},J_{6},J_{8},J_{10}\right)=G_{\text{Tele}}+\epsilon\,G_{\text{Tele},\phi}\delta\phi\nonumber \\
 & \quad+\epsilon^{2}\Big[\frac{1}{2}G_{\text{Tele},\phi\phi}\delta\phi^{2}+G_{\text{Tele},X}X+G_{\text{Tele},T}T+G_{\text{Tele},T_{\text{ax}}}T_{\text{ax}}+G_{\text{Tele},T_{\text{vec}}}T_{\text{vec}}+G_{\text{Tele},I_{2}}I_{2}\Big]+\mathcal{O}(\epsilon^{3})\,,\label{function_expansion}\\
 & G_{j}(\phi,X)=G_{j}+\epsilon\,G_{j,\phi}\,\delta\phi+\epsilon^{2}\,\Big[\frac{1}{2}G_{j,\phi\phi}\delta\phi^{2}+G_{j,X}X\Big]+\mathcal{O}(\epsilon^{3})\,,
\end{align}
where $G_{\text{Tele},i} = G_{\text{Tele},i}(\phi_{0},0,0,0,0,0,0,0,0,0,0,0)$ for $i=\{\phi,\phi\phi,X,T,T_{\text{ax}},T_{\text{vec}},I_{2}\}$ such that $\phi\phi$ subscript stands for the second order derivative with respect to $\phi $ and $G_{j,k} = G_{j,k}(\phi_{0},0)$ for $j=\{2,3,4,5\}$ and $k=\{\phi,\phi\phi,X\}$ are constants. In these expansions, we also note that the $J_i$ contributions do not appear due to their generically higher order nature. Therefore, the linearized field equations $W_{\mu\nu}$, $\widehat{W}$ are obtained by varying the action with respect to $\tau_{\mu\nu}$ and $\delta \phi$ accordingly.
\begin{align}
W_{\mu\nu} & =\left(G_{\text{Tele}}+G_{2}\right)\eta_{\mu\nu}\nonumber \\
 & \quad+\epsilon\,\Big[\left(G_{\text{Tele}}+G_{2}\right)\left(\tau\,\eta_{\mu\nu}-\tau_{\mu\nu}\right)-2G_{\text{Tele},T_{\text{vec}}}\left(\partial^{\lambda}\partial_{\mu}\tau_{\lambda\nu}-\partial_{\mu}\partial_{\nu}\tau-\partial_{\lambda}\partial_{\beta}\tau^{\lambda\beta}\eta_{\mu\nu}+\Box\tau\,\eta_{\mu\nu}\right)\nonumber \\
 & \quad+\left(-2G_{\text{Tele},T}+2G_{4}\right)\left(\Box\tau_{(\mu\nu)}-\partial^{\lambda}\partial_{\mu}\tau_{(\nu\lambda)}-\partial^{\lambda}\partial_{\nu}\tau_{(\mu\lambda)}+\partial_{\mu}\partial_{\nu}\tau+\partial_{\lambda}\partial_{\beta}\tau^{\lambda\beta}\eta_{\mu\nu}-\Box\tau\,\eta_{\mu\nu}\right)\nonumber \\
 & \quad+\frac{4}{9}G_{\text{Tele},T_{\text{ax}}}\left(\Box\tau_{[\mu\nu]}-\partial^{\lambda}\partial_{\nu}\tau_{[\mu\lambda]}+\partial^{\lambda}\partial_{\mu}\tau_{[\nu\lambda]}\right)+\left(-G_{\text{Tele},I_{2}}+2G_{4,\phi}\right)\left(\partial_{\mu}\partial_{\nu}\delta\phi-\eta_{\mu\nu}\Box\delta\phi\right)\Big]\,,\\[4pt]
\widehat{W} & =G_{\text{Tele},\phi}+G_{2,\phi}+\epsilon\,\Big[\left(G_{\text{Tele},\phi}+G_{2,\phi}\right)\,\tau+\left(G_{\text{Tele},I_{2}}-2G_{4,\phi}\right)\left(\Box\tau-\partial_{\lambda}\partial_{\beta}\tau^{\lambda\beta}\right)\nonumber \\
 & \quad+\left(G_{\text{Tele},\phi\phi}+G_{2,\phi\phi}\right)\,\delta\phi+\left(G_{\text{Tele},X}+G_{2,X}-2G_{3,\phi}\right)\Box\delta\phi\Big]\,,
\end{align}
where $\tau:=\eta^{\mu\nu}\tau_{\mu\nu}$ and $\Box \coloneqq \partial_{\lambda} \partial^{\lambda}$ is the d'Alembert operator of Minkowski space. In order to obtain the on-shell perturbed field equations, we impose the background field equations
\begin{eqnarray}
        0 &=& G_{\text{Tele}} + G_{2}\,,\\
        0 &=& G_{\text{Tele},\phi} + G_{2,\phi}\,.
\end{eqnarray}
Thus, the linearised field equations are reduced to
\begin{align}
W_{\mu\nu} & =-2G_{\text{Tele},T_{\text{vec}}}\left(\partial^{\lambda}\partial_{\mu}\tau_{\lambda\nu}-\partial_{\mu}\partial_{\nu}\tau-\partial_{\lambda}\partial_{\sigma}\tau^{\lambda\sigma}\eta_{\mu\nu}+\Box\tau\,\eta_{\mu\nu}\right)\nonumber \\
 & \quad+\left(-2G_{\text{Tele},T}+2G_{4}\right)\left(\Box\tau_{(\mu\nu)}-\partial^{\lambda}\partial_{\mu}\tau_{(\nu\lambda)}-\partial^{\lambda}\partial_{\nu}\tau_{(\mu\lambda)}+\partial_{\mu}\partial_{\nu}\tau+\partial_{\lambda}\partial_{\sigma}\tau^{\lambda\sigma}\eta_{\mu\nu}-\Box\tau\,\eta_{\mu\nu}\right)\nonumber \\
 & \quad+\frac{4}{9}G_{\text{Tele},T_{\text{ax}}}\left(\Box\tau_{[\mu\nu]}-\partial^{\lambda}\partial_{\nu}\tau_{[\mu\lambda]}+\partial^{\lambda}\partial_{\mu}\tau_{[\nu\lambda]}\right)+\left(-G_{\text{Tele},I_{2}}+2G_{4,\phi}\right)\left(\partial_{\mu}\partial_{\nu}\delta\phi-\eta_{\mu\nu}\Box\delta\phi\right)\,,\label{Wmunutotal}\\[4pt]
\widehat{W} & =\left(G_{\text{Tele},I_{2}}-2G_{4,\phi}\right)\left(\Box\tau-\partial_{\lambda}\partial_{\sigma}\tau^{\lambda\sigma}\right)+\left(G_{\text{Tele},\phi\phi}+G_{2,\phi\phi}\right)\,\delta\phi+\left(G_{\text{Tele},X}+G_{2,X}-2G_{3,\phi}\right)\Box\delta\phi\,.\label{Wscalartotal}
\end{align}

Here, Eq.~(\ref{Wmunutotal}) can be decomposed into symmetric and antisymmetric parts as follows
\begin{align}
W_{(\mu\nu)} & =-2G_{\text{Tele},T_{\text{vec}}}\left(\partial^{\lambda}\partial_{(\mu}\tau_{|\lambda|\nu)}-\partial_{\mu}\partial_{\nu}\tau-\partial_{\lambda}\partial_{\beta}\tau^{\lambda\beta}\eta_{\mu\nu}+\Box\tau\eta_{\mu\nu}\right)+\left(-G_{\text{Tele},I_{2}}+2G_{4,\phi}\right)\left(\partial_{\mu}\partial_{\nu}\delta\phi-\eta_{\mu\nu}\Box\delta\phi\right)\nonumber \\
 & \quad+2\left(-G_{\text{Tele},T}+G_{4}\right)\left(\Box\tau_{(\mu\nu)}-\partial^{\lambda}\partial_{\mu}\tau_{(\nu\lambda)}-\partial^{\lambda}\partial_{\nu}\tau_{(\mu\lambda)}+\partial_{\mu}\partial_{\nu}\tau+\partial_{\lambda}\partial_{\beta}\tau^{\lambda\beta}\eta_{\mu\nu}-\Box\tau\eta_{\mu\nu}\right)\,,\label{eq:FE_symm}\\[6pt]
W_{[\mu\nu]} & =-2G_{\text{Tele},T_{\text{vec}}}\partial^{\lambda}\partial_{[\mu}\tau_{|\lambda|\nu]}+\frac{4}{9}G_{\text{Tele},T_{\text{ax}}}\left[\Box\tau_{[\mu\nu]}-\partial^{\lambda}\partial_{\nu}\tau_{[\mu\lambda]}+\partial^{\lambda}\partial_{\mu}\tau_{[\nu\lambda]}\right]\,,\label{eq:FE_antisymm}
\end{align}
where the antisymmetric part coincides with the field equations one would obtain by varying with respect to a non-trivial inertial spin connection. Using the symmetry operators gives a more workable formulation to the perturbation equations which will then be studied in the sections that follow.

\section{Scalar--Vector--Tensor Decomposition and Propagating Degrees of Freedom Analysis}\label{sec:dof_analysis}

Following the notation of Ref.~\cite{Bahamonde:2020lsm}, the tetrad perturbation
$\delta e^{A}{}_{\mu}$, in Minkowski space, can be split as follows
\begin{equation}
    \delta e^{A}{}_{\mu}:=\left[\arraycolsep=3pt\def\arraystretch{1.4}\begin{array}{cc} -\varphi & -\left(\partial_{i}\beta+\beta_{i}\right)\\
    \delta^{I}{}_{i}\left(\partial^{i}b+b^{i}\right) & \delta^{Ii}\left(-\psi\delta_{ij}+\partial_{i}\partial_{j}h+2\partial_{(i}h_{j)}+\frac{1}{2}h_{ij}+\epsilon_{ijk}\left(\partial^{k}\sigma+\sigma^{k}\right)\right)
\end{array}\right],\label{eq:tetrad_perturbation}
\end{equation}
which will further generate $\tau_{\mu\nu}$, using Eq.~(\ref{eq:taumunu}), to give
\begin{equation}
    \tau_{\mu\nu}=\left[\arraycolsep=3pt\def\arraystretch{1.4}\begin{array}{cc} -\varphi & -\left(\partial_{i}\beta+\beta_{i}\right)\\
    (\partial_{i}b+b_{i}) & \left(-\psi\delta_{ij}+\partial_{i}\partial_{j}h+2\partial_{(i}h_{j)}+\frac{1}{2}h_{ij}+\epsilon_{ijk}\left(\partial^{k}\sigma+\sigma^{k}\right)\right)
\end{array}\right],\label{eq:deltatetradgreek_perturbation}
\end{equation}
and one can also produce the perturbation of the metric perturbation since $\delta g_{\mu\nu}=2\tau_{(\mu\nu)}$, and so
\begin{equation}
    \delta g_{\mu\nu}=\left[\arraycolsep=3pt\def\arraystretch{1.4}\begin{array}{cc} - 2\,\varphi & \left(\partial_{i}\mathcal{B}+\mathcal{B}_{i}\right)\\
    \left(\partial_{i}\mathcal{B}+\mathcal{B}_{i}\right) & 2\left(-\psi\delta_{ij}+\partial_{i}\partial_{j}h+2\partial_{(i}h_{j)}+\frac{1}{2}h_{ij}\right)
\end{array}\right],\label{eq:delta_metric_perturbation}
\end{equation}
where $\left\{ \varphi,\beta,b,\psi,h\right\} $ are scalars with
$\sigma$ being a pseudo scalar, $\left\{ \beta_{i},b_{i},h_{i}\right\} $
are vectors with $\sigma^{k}$ being a pseudo vector and $h_{ij}$
the tensor modes, and which represent the regular perturbation of the metric tensor about Minkowski space. All indexed quantities are divergence-less, meaning that $\partial^{i}X_{ijk..l}=0$, $h_{ij}=h_{(ij)}$ and $\delta^{ij}h_{ij}=0$.
We also have denoted $\mathcal{B}:=b-\beta$ and $\mathcal{B}_{i}:=b_{i}-\beta_{i}$
to stress the fact that in the perturbation of the metric Eq.~(\ref{eq:delta_metric_perturbation})
the off diagonal elements are a single scalar and vector. While Latin indixes are used for inertial coordinates, we reserve the middle range Latin indices $I,J,K,\dots$ and $i,j,k,\dots$ for purely spacial coordinates in the spacial inner bundle and the spacial spacetime manifold respectively.\medskip

The gauge transformation of $\delta e^{A}{}_{\nu}$, under a linear change of coordinates described by $\widetilde{x}^{\mu}\rightarrow x^{\mu}+\xi^{\mu}$, can be expressed as
\begin{align}
\widetilde{\delta e}^{A}{}_{\mu} & \rightarrow\delta e{}^{A}{}_{\mu}+\mathcal{L}_{\xi}e{}^{A}{}_{\mu}.\label{eq:pert_tetr_gauge-1-1}
\end{align}
where $\mathcal{L}_{\xi}$ is the Lie derivative along the flow of $\xi^\mu$. We further split $\xi^{\mu}$ as $\xi^{\mu}=\left\{ \xi^{0},\omega(\xi^{i}+\delta{}^{ij}\partial{}_{j}\xi)\right\} $
where $\partial_{i}\xi^{i}=0$.
One can then unpack Eq.~(\ref{eq:pert_tetr_gauge-1-1}) to all its individual components as
\begin{subequations}\label{eq:gauge_transformations}\begin{eqnarray}
    \widetilde{\varphi} & =& \varphi-\dot{\xi}{}^{0}\,,\quad \widetilde{\psi} = \psi\,,\quad \widetilde{\beta}  =  \beta-\xi{}^{0}\,,\quad \widetilde{\beta_{i}} =  \beta_{i}\,,\\
    \widetilde{b}  &=&  b-\dot{\xi}\,, \quad \widetilde{b}_{i} = b_{i}+\dot{\xi_{i}}\,,\quad  \widetilde{\sigma} = \sigma\,,\quad \widetilde{\sigma}_{i} = \sigma^{i}-\frac{1}{2}\epsilon^{i}{}_{jk}\partial^{j}\xi^{k}\,,\\
    \widetilde{h} & =&  h-\xi\,,\quad \widetilde{h}_{i} =  h_{i}+\frac{1}{2}\xi_{i}\,,\quad  \widetilde{h}_{ij} = h_{ij}\,.
\end{eqnarray}\end{subequations}
As it is evident $\psi, \sigma$, $\beta_i$ and $h_{ij}$ are already gauge invariant in the Minkowski background. In addition, $\delta \phi$ is also  gauge invariant in this setting since its background value is constant $\phi(t) \equiv \phi_0$. In general one can built a few other gauge invariant quantities from Eq.~(\ref{eq:gauge_transformations}) according to their similarities in the gauge transformations as
\begin{align}
\chi & :=b-\dot{h},\label{eq:gauge_inv1}\\
\Phi & :=\varphi-\dot{\beta},\\
\Sigma_{j} & :=h_{j}k{}^{2}+i\varepsilon{}_{jlp}k{}^{l}\sigma{}^{p},\\
\Xi_{j} & :=ik^{2}b{}_{j}-2\varepsilon{}_{jlp}k{}^{l}\dot{\sigma}{}^{p},\\
\Lambda_{j} & :=-b_{j}+2\dot{h}{}_{j}.\label{eq:gauge_inv2}
\end{align}
We raise and lower the indices as $X_{0}=X^{0}$, $X_{j}=-X^{j}$ and use the conventions $\Box:=\partial_{\mu}\partial^{\mu}=\partial_{0}^{2}-\partial^{2}$,
where the spacial Laplacian is defined as $\partial^{2}:=-\eta^{ij}\partial_{j}\partial_{i}=\delta^{ij}\partial_{i}\partial_{j}$.
Also we decompose the perturbations in Fourier space as $f(x^{\mu})\rightarrow f(k^{\mu})e^{-i\omega t+ik_{j}x^{j}}$ and we define the norm of the wave
covector as $k^{2}:=-\eta^{ij}k_{i}k_{j}=\delta^{ij}k_{i}k_{j}$. \medskip

We split the field equations in Eqs.~(\ref{Wmunutotal}--\ref{Wscalartotal})
in a SVT manner by inserting Eq.~(\ref{eq:deltatetradgreek_perturbation})
and the we switch to purely gauge invariant variables. This is realised by using the gauge invariant variables $\left(\Phi,\chi,\Sigma{}_{i},\Lambda{}_{i}\right)$ introduced in Eqs.~(\ref{eq:gauge_inv1}-\ref{eq:gauge_inv2}). The scalar sector consists of 5 linearly independent equations (considering no possible soft-properties extension \cite{Saridakis:2021qxb}) for the corresponding gauge invariant scalars $\left(\delta\phi,\psi,\Phi,\chi,\sigma\right)$ which are given as
\begin{align}
W_{00} & =k^{2}\Big((G_{\text{Tele},I_{2}}-2G_{4,\phi})\delta\phi+2G_{\text{Tele},T_{\text{vec}}}\Phi-4(G_{4}-G_{\text{Tele},T}+G_{\text{Tele},T_{\text{vec}}})\psi\Big)\,,\label{eq:Scalareq1}\\[4pt]
k^{j}W{}_{j0} & =ik^{2}\bigl((G_{\text{Tele},I_{2}}-2G_{4,\phi})\delta\dot{\phi}+2G_{\text{Tele},T_{\text{vec}}}\chi k{}^{2}-2(2(G_{4}-G_{\text{Tele},T})+3G_{\text{Tele},T_{\text{vec}}})\dot{\psi}\bigr)\,,\label{eq:Scalareq2}\\[4pt]
\eta_{jl}W{}^{jl} & =3(G_{\text{Tele},I_{2}}-2G_{4,\phi})\delta\ddot{\phi}+2k{}^{2}\Big((2(G_{4}-G_{\text{Tele},T})+3G_{\text{Tele},T_{\text{vec}}})\dot{\chi}+(G_{\text{Tele},I_{2}}-2G_{4,\phi})\delta\phi\nonumber \\
 &  \,\,\,\,\,+2(G_{4}-G_{\text{Tele},T}+G_{\text{Tele},T_{\text{vec}}})\Phi-2(G_{4}-G_{\text{Tele},T}+2G_{\text{Tele},T_{\text{vec}}})\psi\Big)\nonumber \\
 &  \,\,\,\,\, -6\Big(2(G_{4}-G_{\text{Tele},T})+3G_{\text{Tele},T_{\text{vec}}}\Big)\ddot{\psi}\,,\label{eq:Scalareq3}\\[4pt]
k_{l}\epsilon^{ljk}W_{jk} & =-\frac{8}{3}iG_{\text{Tele},T_{\text{ax}}}k{}^{2}\bigl(\ddot{\sigma}+k{}^{2}\sigma\bigr)\,,\label{eq:Scalareq4}\\[4pt]
\widehat{W} & =(G_{\text{Tele},X}+G_{2,X}-2G_{3,\phi})\delta\ddot{\phi}+\delta\phi\Big(G_{\text{Tele},\phi\phi}+G_{2,\phi\phi}+(G_{\text{Tele},X}+G_{2,X}-2G_{3,\phi})k{}^{2}\Big)\nonumber \\
 &  \,\,\,\,\, +(G_{\text{Tele},I_{2}}-2G_{4,\phi})\bigl(-k{}^{2}(\dot{\chi}+\Phi-2\psi)+3\ddot{\psi}\bigr)\,.\label{eq:Scalareq5}
\end{align}
On the other hand, the vector part consists of 3 linearly independent equations for the gauge invariant variables $\left(\beta{}_{i},\Sigma{}_{i},\Lambda{}_{i}\right)$
\begin{align}
W_{0j} & =\frac{1}{9}\Big\{18G_{\text{Tele},T_{\text{vec}}}\ddot{\beta}{}_{j}+k{}^{2}\Big[\Big(2G_{\text{Tele},T_{\text{ax}}}-9(G_{4}-G_{\text{Tele},T})\Big)\Lambda{}_{j}-\Big(9(G_{4}-G_{\text{Tele},T})+2G_{\text{Tele},T_{\text{ax}}}\Big)\beta{}_{j}\Big]\nonumber \\
 &  \,\,\,\,\,-2\Big(9G_{\text{Tele},T_{\text{vec}}}+2G_{\text{Tele},T_{\text{ax}}}\Big)\dot{\Sigma}{}_{j}\Big\}\,,\label{eq:Vectoreq1}\\[4pt]
W_{j0} & =\frac{1}{9}\Big\{ k{}^{2}\Big[\Big(2G_{\text{Tele},T_{\text{ax}}}-9(G_{4}-G_{\text{Tele},T})\Big)\beta{}_{j}-\Big(9(G_{4}-G_{\text{Tele},T})+2G_{\text{Tele},T_{\text{ax}}}\Big)\Lambda{}_{j}\Big]+4G_{\text{Tele},T_{\text{ax}}}\dot{\Sigma}{}_{j}\Big\}\,,\label{eq:Vectoreq2}\\[4pt]
k^{l}W{}_{lj} & =-\frac{1}{9}i\Big\{ k{}^{2}\Big[\Big(9(G_{4}-G_{\text{Tele},T})+2(9G_{\text{Tele},T_{\text{vec}}}+G_{\text{Tele},T_{\text{ax}}})\Big)\dot{\beta}{}_{j}+\Big(9(G_{4}-G_{\text{Tele},T})-2G_{\text{Tele},T_{\text{ax}}}\Big)\dot{\Lambda}{}_{j}\nonumber \\
 & \,\,\,\,\,-18G_{\text{Tele},T_{\text{vec}}}\Sigma{}_{j}\Big]+4G_{\text{Tele},T_{\text{ax}}}\ddot{\Sigma}{}_{j}\Big\}\,,\label{eq:Vectoreq3}
\end{align}
while the tensor field is described by one equation
\begin{align}
    W_{ij} & =(G_{4}-G_{\text{Tele},T})(\ddot{h}_{ij}+k{}^{2}h_{ij})\,.\label{eq:Tensoreq}
\end{align}

In order to solve the full system of equations for the scalar, vector and tensor sector we rewrite the corresponding equations as matrices in the following way:\medskip

\underline{\textbf{Scalar sector}}:
\begin{equation}
   \left(\arraycolsep=2pt\def\arraystretch{1.3}\begin{array}{ccccc}
M_{S11} & M_{S12} & M_{S13} & 0 & 0\\
M_{S21} & M_{S22} & 0 & M_{S24} & 0\\
M_{S31} & M_{S32}& M_{S33} & M_{S34} & 0\\
0 & 0 & 0 & 0 & M_{S45}\\
M_{S51} & M_{S52} & M_{S53} & M_{S54} & 0
\end{array}\right)\left(\arraycolsep=2pt\def\arraystretch{1.3}\begin{array}{c}
\delta\phi\\
\psi\\
\Phi\\
\sigma\\
\chi
\end{array}\right)=:M_S Y_S=0\,,\label{eq:ScalarsMat}
\end{equation}
where the components of the matrix $M_S$ are
\begin{subequations}
    \begin{align}
M_{S11} & =(G_{\text{Tele},I_{2}}-2G_{4,\phi})k{}^{2}\,,\quad M_{S12}=-4(G_{4}-G_{\text{Tele},T}+G_{\text{Tele},T_{\text{vec}}})k{}^{2}\,,\quad M_{S13}=M_{S12}-\frac{M_{S22}}{\omega}\,,\\
M_{S21} & =\omega M_{S11}\,,\quad M_{S22}=-2(2(G_{4}-G_{\text{Tele},T})+3G_{\text{Tele},T_{\text{vec}}})\omega k{}^{2}\,,\quad M_{S24}=\frac{ik^{2}}{\omega}(\omega M_{S12}-M_{S22})\,,\\
M_{S31} & =\bigl(2-\frac{3\omega^{2}}{k^{2}}\bigr)M_{S11}\,,\quad M_{S32}=-M_{S12}+\frac{M_{S22}}{\omega}\bigl(2-\frac{3\omega^{2}}{k^{2}}\bigr)\,,\quad M_{S33}=-M_{S12}\,,\\
M_{S34} & =iM_{S22}\,,\quad M_{S45}=\frac{8}{3}i\,G_{\text{Tele},T_{\text{ax}}}k^{2}(\omega^{2}-k^{2})\,,\\
M_{S51} & =G_{\text{Tele},\phi\phi}+G_{2,\phi\phi}-(G_{\text{Tele},X}+G_{2,X}-2G_{3,\phi})(\omega^{2}-k^{2})\,,\quad M_{S52}=\bigl(2-\frac{3\omega^{2}}{k^{2}}\bigr)M_{S11}\,,\\
M_{S53} & =-M_{S11}\,,\quad M_{S54}=i\omega M_{S11}\,.
\end{align}
\end{subequations}
where we have expressed all the matrix elements wrt to the unique elements $M_{S11},M_{S12},M_{S22}$ and $M_{S51}$ whenever possible.

\normalsize \underline{\textbf{Vector sector}}:
\begin{equation}
\left(\arraycolsep=2pt\def\arraystretch{1.3}\begin{array}{ccc}
M_{V11} & M_{V12} & M_{V13}\\
M_{V21} & M_{V22} & M_{V23}\\
M_{V31} & M_{V32} & M_{V33}
\end{array}\right)\left(\arraycolsep=2pt\def\arraystretch{1.3}\begin{array}{c}
\beta{}_{j}\\
\Sigma{}_{j}\\
\Lambda{}_{j}
\end{array}\right)=:M_{V}Y_{V}=0\,,\label{eq:VectorsMat}
\end{equation} where the components of the vectorial matrix $M_V$ are
\begin{subequations}
    \begin{align}
M_{V11} & =-2G_{\text{Tele},T_{\text{vec}}}\omega^{2}-\tfrac{1}{9}(9G_{4}-G_{\text{Tele},T}+2G_{\text{Tele},T_{\text{ax}}})k{}^{2}\,,\\
M_{V12} & =\frac{i}{k^{2}}(\omega M_{V13}-M_{V31})\,,\quad M_{V13}=\tfrac{1}{9}(-9G_{4}-G_{\text{Tele},T}+2G_{\text{Tele},T_{\text{ax}}})k{}^{2}\,,\\
M_{V21} & =M_{V13}\,,\quad M_{V22}=\frac{-i\omega}{k^{2}(\omega^{2}-k{}^{2})}\Bigl(k{}^{2}M_{V11}+\bigl(\omega^{2}-k{}^{2}\bigr)M_{V13}-\omega M_{V31}\Bigr)\,,\\
M_{V23} & =\frac{1}{\omega^{2}-k{}^{2}}\bigl(-k{}^{2}M_{V11}+\omega M_{V31}\bigr)\,,\\
M_{V31} & =-\tfrac{1}{9}\bigl(9G_{4}-G_{\text{Tele},T}+2(9G_{\text{Tele},T_{\text{vec}}}+G_{\text{Tele},T_{\text{ax}}})\bigr)\omega k{}^{2}\,,\\
M_{V32} & =\frac{i}{\omega k{}^{2}}\Bigl(\omega(k{}^{2}M_{V11}+\omega^{2}M_{V13})-\bigl(\omega^{2}+k{}^{2}\bigr)M_{V31}\Bigr)\,,\quad M_{V33}=\omega M_{V13}\,,
\end{align}
\end{subequations}
where we have expressed all the matrix elements wrt to the unique elements $M_{V11},M_{V13}$ and $M_{V31}$ whenever possible.

\normalsize \underline{\textbf{Tensor sector}}:
\begin{align}
-(G_{4}-G_{\text{Tele},T})(\omega^{2}-k^{2})=:M_T Y_T=0\,.\label{eq:TensorMat}
\end{align}
\noindent where we stress that $G_{4}-G_{\text{Tele},T}\neq0$. An interesting feature of this representation is that all three sector equations can be collected into one master in a block diagonal manner
\begin{align}
\left(\arraycolsep=2pt\def\arraystretch{1.3}\begin{array}{ccc}
M_{S} & 0 & 0\\
0 & M_{V} & 0\\
0 & 0 & M_{T}
\end{array}\right)\left(\arraycolsep=2pt\def\arraystretch{1.3}\begin{array}{c}
Y_{S}\\
Y_{V}\\
Y_{T}
\end{array}\right) &=:MY =0\,,\label{eq:TotalMat}
\end{align}
where $M_{V}$ and $M_{T}$ are to be understood as $M_{V}\otimes I_{2}$
and $M_{T}\otimes I_{2}=M_{T}I_{2}$, and $I_{2}$ being the identity
matrix of dimension 2. This is to account for the fact that all indexed
fields carry 2 components, i.e, $\beta{}_{j}=\left(\beta{}_{1},\beta{}_{2}\right)$. The propagation of GWs can analogously be characterized by a principal polynomial \cite{Hohmann:2018jso,hormander2015analysis,hormander2015analysisII} of the perturbation equations which vanishes. Thus, we can calculate the principal polynomial $P(k):=\det(M)=\det(M_{S})\det(M_{V})\det(M_{T})$ of $M$ as
\begin{align}
P(k) & =-2^{14}3^{-5}(G_{4}-G_{\text{Tele},T}){}^{5}G_{\text{Tele},T_{\text{vec}}}{}^{3}G_{\text{Tele},T_{\text{ax}}}{}^{3}k{}^{12}\bigl(\omega^{2}-k{}^{2}\bigr)^{8}\Bigl(\tilde{c}_{1}+\tilde{c}_{2}\bigl(\omega^{2}-k{}^{2}\bigr)\Bigr)\,,\label{eq:Pk}
\end{align}
where
\begin{align}
\tilde{c}_{1} & :=2(G_{\text{Tele},\phi\phi}+G_{2,\phi\phi})(2(G_{4}-G_{\text{Tele},T})+3G_{\text{Tele},T_{\text{vec}}}),\label{eq:c1tdef}\\
\tilde{c}_{2} & :=-3(G_{\text{Tele},I_{2}}-2G_{4,\phi})^{2}-2(G_{\text{Tele},X}+G_{2,X}-2G_{3,\phi})(2(G_{4}-G_{\text{Tele},T})+3G_{\text{Tele},T_{\text{vec}}})\,.\label{eq:c2tdef}
\end{align}
it is also convenient to define the quantities
\begin{align}
    \tilde{c}_{3} & :=-G_{\text{Tele},\phi\phi}-G_{2,\phi\phi}\,,\label{eq:c3tdef}\\
    \tilde{c}_{4} & :=G_{\text{Tele},X}+G_{2,X}-2G_{3,\phi}\,,\label{eq:c4tdef}
\end{align}
and
\begin{align}
Z_{1} & :=-\frac{(G_{4}-G_{\text{Tele},T})G_{2,\phi\phi}}{-3G_{4,\phi}{}^{2}+(G_{4}-G_{\text{Tele},T})(2G_{3,\phi}-G_{2,X})}\,,\label{Z1}\\
\nonumber \\
Z_{2} & :=\frac{\bigl(3(G_{\text{Tele},I_{2}}-2G_{4,\phi})^{2}+2(2(G_{4}-G_{\text{Tele},T})+3(G_{\text{Tele},T_{\text{vec}}}))(G_{\text{Tele},X}+G_{2,X})\bigr)}{4(2(G_{4}-G_{\text{Tele},T})+3(G_{\text{Tele},T_{\text{vec}}}))}\,.\label{Z2}
\end{align}
Given the form of $P(k)$ in Eq.~(\ref{eq:Pk}), we expect that in some branches of the theory there exist massive or massless modes or even both. A detailed analysis needs to be carried out for all the cases where $P(k)$ is degenerate, i.e $P(k)=0$ for any $k$. One of these cases can be generated by $G_{\text{Tele},T_{\text{vec}}}=0$ for example. Thus, in general this is an iterative process in order to find the reduced and non-degenerate instances of $P(k)$ and then find the corresponding dispersion relations $\omega=\omega(k)$. Hence after obtaining the dispersion relations that correspond to non-degenerate $P(k)$ we obtain the solutions of the system in Eq.~(\ref{eq:TotalMat}), all of which span a null space (the space of all solutions of the system of Eq.~(\ref{eq:TotalMat})).\medskip
\begin{center}
\begin{table}[htp!]
\begin{centering}
\setlength{\tabcolsep}{7pt} \renewcommand{\arraystretch}{1.5}{%
\resizebox{\columnwidth}{!}{\begin{tabular}{|c|l|c|c|c||c|c|c|}
\hline
\multirow{3}{*}{\textbf{Cases}} & \multirow{3}{*}{\textbf{Conditions}} & \multicolumn{5}{c|}{\textbf{Sectors}} & \multirow{3}{*}{\textbf{PDoF}}\tabularnewline
\cline{3-7} \cline{4-7} \cline{5-7} \cline{6-7} \cline{7-7}
 &  & \multicolumn{3}{c||}{Massless$\left(\omega^{2}=k{}^{2}\right)$} & \multicolumn{2}{c|}{Massive$\left(\omega^{2}-k{}^{2}=m^{2}\right)$} & \tabularnewline
\cline{3-7} \cline{4-7} \cline{5-7} \cline{6-7} \cline{7-7}
 &  & Scalar & Vector & \multicolumn{1}{c||}{Tensor} & Scalar & $m^{2}$ & \tabularnewline
\hline
\multirow{1}{*}{--} & $G_{4}-G_{\text{Tele},T}\neq0$ & \multirow{1}{*}{-} & \multirow{1}{*}{-} & \multirow{1}{*}{1} & \multirow{1}{*}{-} & \multirow{1}{*}{-} & \multirow{1}{*}{2}\tabularnewline
\hline
\multirow{2}{*}{0} & $G_{\text{Tele},T_{\text{vec}}}=0,\,G_{\text{Tele},T_{\text{ax}}}=0,$  & \multicolumn{3}{c||}{} &  &  & \tabularnewline
 & $G_{\text{Tele},I_{2}}=0,G_{\text{Tele},X}=0,G_{\text{Tele},\phi\phi}=0$ & \multicolumn{3}{c||}{} &  &  & \tabularnewline
\hline
\multirow{2}{*}{0.I} & $G_{2,\phi\phi}\neq0$ and  & \multirow{2}{*}{-} & \multirow{2}{*}{-} & \multirow{2}{*}{1} & \multirow{2}{*}{1} & \multirow{2}{*}{$Z_1$} & \multirow{2}{*}{3}\tabularnewline
 &
 $-3G_{4,\phi}{}^{2}+(G_{4}-G_{\text{Tele},T})\left(2G_{3,\phi}-G_{2,X}\right)\neq0$&  &  &  &  &  & \tabularnewline
\hline
\multirow{2}{*}{0.II} & $G_{2,\phi\phi}=0$ and& \multirow{2}{*}{-} & \multirow{2}{*}{-} & \multirow{2}{*}{1} & \multirow{2}{*}{-} & \multirow{2}{*}{-} & \multirow{2}{*}{2}\tabularnewline
& $-3G_{4,\phi}{}^{2}+(G_{4}-G_{\text{Tele},T})\left(2G_{3,\phi}-G_{2,X}\right)=0$ &  &  &  &  &  & \tabularnewline
\hline
\multirow{2}{*}{1} & $G_{\text{Tele},T_{\text{vec}}}\neq0,\,G_{\text{Tele},T_{\text{ax}}}\neq0$ & \multirow{2}{*}{2} & \multirow{2}{*}{1} & \multirow{2}{*}{1} & \multirow{2}{*}{1} & \multirow{2}{*}{$-\tilde{c}_{1}/\tilde{c}_{2}$} & \multirow{2}{*}{7}\tabularnewline
 &$\tilde{c}_{1}\neq0,\tilde{c}_{2}\neq0$
 &  &  &  &  &  & \tabularnewline
\hline
2 & $G_{\text{Tele},T_{\text{vec}}}\neq0,\, G_{\text{Tele},T_{\text{ax}}}\neq0,\tilde{c}_{1}=0,\tilde{c}_{2}=0$ & \multicolumn{3}{c||}{} & \multirow{2}{*}{} & \multirow{2}{*}{} & \multirow{2}{*}{}\tabularnewline
\cline{1-2} \cline{2-2}
2.I & $G_{\text{Tele},T_{\text{ax}}}\neq0,\tilde{c}_{3}\neq0,\tilde{c}_{4}\neq0$ & \multicolumn{3}{c||}{} &  &  & \tabularnewline
\hline
2.I.a & $G_{\text{Tele},T_{\text{ax}}}\neq0,\tilde{c}_{3}\neq0,\tilde{c}_{4}\neq0$ & 1 & 1 & 1 & 1 & $-\tilde{c}_{3}/\tilde{c}_{4}$ & 6\tabularnewline
\hline
2.I.b & $
G_{\text{Tele},T_{\text{ax}}}\neq0,\tilde{c}_{3}=0,\tilde{c}_{4}=0
$ & 1 & 1 & 1 & - & - & 5\tabularnewline
\hline
2.II & $G_{3,\phi}=Z_2,\,G_{2,\phi\phi}=-G_{\text{Tele},\phi\phi}$ & \multicolumn{3}{c||}{} &  &  & \tabularnewline
\hline
2.II.a & $G_{\text{Tele},I_{2}}-2G_{4,\phi}\neq0$ & 2 & 1 & 1 & - & - & 6\tabularnewline
\hline
2.II.b & $G_{\text{Tele},I_{2}}-2G_{4,\phi}=0$ & \multirow{1}{*}{1} & 1 & 1 & - & - & 5\tabularnewline
\hline
3 & $G_{\text{Tele},T_{\text{vec}}}\neq0,\, G_{\text{Tele},T_{\text{ax}}}=0,\tilde{c}_{1}\neq0,\tilde{c}_{2}\neq0$ & 1 & - & 1 & 1 & $-\tilde{c}_{1}/\tilde{c}_{2}$ & 4\tabularnewline
\hline
4 & $G_{\text{Tele},T_{\text{vec}}}\neq0,\,G_{\text{Tele},T_{\text{ax}}}=0,\tilde{c}_{1}=0,\tilde{c}_{2}=0$ & \multicolumn{3}{c||}{} & \multirow{2}{*}{} & \multirow{2}{*}{} & \multirow{2}{*}{}\tabularnewline
\cline{1-2} \cline{2-2}
4.I & $G_{\text{Tele},T_{\text{vec}}}=-\tfrac{2}{3}(G_{4}-G_{\text{Tele},T}),\,G_{4,\phi}=\tfrac{1}{2}G_{\text{Tele},I_{2}}$ & \multicolumn{3}{c||}{} &  &  & \tabularnewline
\hline
4.I.a & $\tilde{c}_{3}\neq0,\tilde{c}_{4}\neq0$ & - & - & 1 & 1 & $-\tilde{c}_{3}/\tilde{c}_{4}$ & 3\tabularnewline
\hline
4.I.b & $\tilde{c}_{3}=0,\tilde{c}_{4}=0$ & - & - & 1 & - & - & 2\tabularnewline
\hline
4.II & $G_{3,\phi}=Z_2,\,G_{2,\phi\phi}=-G_{\text{Tele},\phi\phi}$ & \multicolumn{3}{c||}{} &  &  & \tabularnewline
\hline
4.II.a & $G_{\text{Tele},I_{2}}-2G_{4,\phi}\neq0$ & 1 & - & 1 & - & - & 3\tabularnewline
\hline
4.II.b & $G_{\text{Tele},I_{2}}-2G_{4,\phi}=0$ & - & - & 1 & - & - & 2\tabularnewline
\hline
5 & $G_{\text{Tele},T_{\text{vec}}}=0,\,G_{\text{Tele},T_{\text{ax}}}\neq0,\tilde{c}_{1}\neq0,\tilde{c}_{2}\neq0$ & 1 & - & 1 & 1 & $-\tilde{c}_{1}/\tilde{c}_{2}$ & 4\tabularnewline
\hline
6 & $G_{\text{Tele},T_{\text{vec}}}=0,G_{\text{Tele},T_{\text{ax}}}\neq0,\tilde{c}_{1}=0,\tilde{c}_{2}=0$ & 1 & - & 1 & - & - & 3\tabularnewline
\hline
7 & $G_{\text{Tele},T_{\text{vec}}}=0,\, G_{\text{Tele},T_{\text{ax}}}=0,\,\tilde{c}_{1}\neq0,\tilde{c}_{2}\neq0$ & - & - & 1 & 1 & $-\tilde{c}_{1}/\tilde{c}_{2}$ & 3 \tabularnewline
\hline
8 & $G_{\text{Tele},T_{\text{vec}}}=0,\,G_{\text{Tele},T_{\text{ax}}}=0,\tilde{c}_{1}=0,\tilde{c}_{2}=0$ & - & - & 1 & - & - & 2\tabularnewline
\hline
\end{tabular}}}
\par\end{centering}
\caption{All Branches of the theory are represented with their respective PDoF. To each of the scalar, vector and tensor components correspond 1,2 and 2 DoF respectively. The quantities $\tilde{c}_{i}$ are defined in~\eqref{eq:c1tdef},~\eqref{eq:c2tdef},~\eqref{eq:c3tdef} and \eqref{eq:c4tdef} while $Z_1$ and $Z_2$ are defined in Eqs.~\eqref{Z1}-\eqref{Z2}) and , respectively.}
\label{tab:GW_DoF}
\end{table}
\end{center}
\vspace{-1cm} \hspace{6pt}
It is instructive to go through the simplest cases of GR and $f(T)$ gravity for which only the combination $(G_{4}-G_{\text{Tele},T})\neq0$ in Eq.~(\ref{eq:TotalMat}) to illustrate the core steps of the procedure. It is straightforward to see that after reducing the system to only linearly dependent equations, we obtain that the system with $M_s,M_V$ and $M_T$ being defined in~\eqref{eq:ScalarsMat},~\eqref{eq:VectorsMat} and \eqref{eq:TensorMat}, respectively, becomes
\begin{equation}
  \left(\arraycolsep=2pt\def\arraystretch{1.5}\begin{array}{cccccc}
   M_{11} & 0 & 0 & 0 & 0 & 0\\
   M_{21} & -M_{11} & 0 & 0 & 0 & 0\\
    0 & 0 & \frac{M_{11}}{4}& 0 & 0 & 0\\
    0 & 0 & 0 & \frac{M_{11}}{4} & 0 & 0\\
    0 & 0 & 0 & 0 & M_{55} & 0\\
    0 & 0 & 0 & 0 & 0 & M_{55}
    \end{array}\right)\left(\arraycolsep=2pt\def\arraystretch{1.5}\begin{array}{c}
    \psi\\
    \Phi\\
    \beta{}_{i}\\
    h_{ij}
    \end{array}\right)=MY\,,
\end{equation}
with the components $M_i$ being
\begin{align}
    M_{11}&= -4(G_{4}-G_{\text{Tele},T})k{}^{2}\,,\quad  M_{21}= 4(G_{4}-G_{\text{Tele},T})\bigl(3\omega^{2}-k{}^{2}\bigr)\,,\\
 M_{55}&=(G_{4}-G_{\text{Tele},T})\bigl(-\omega^{2}+k{}^{2}\bigr)\,.
\end{align}
\normalsize
from which it is already obvious that only the tensor modes propagate,
nevertheless we will go through the steps of the standard procedure used throughout the cases. The principal polynomial
is $P(k)=-16(G_{4}-G_{\text{Tele},T}){}^{6}k{}^{8}\bigl(\omega^{2}-k{}^{2}\bigr)^{2}$,
which is not degenerate, and thus our dispersion relation describes
the usual massless speed of light propagation $\omega^{2}-k{}^{2}=0$.
Substituting in the dispersion relation into Eq.~(\ref{eq:TotalMat})
and solving the system, we generate the null space of $M$ i.e. all
the space of all linearly independent solutions that the system can
have. We then calculate the general solution $Y_{\left|k\right|}$
which is a linear combination, that uses as coefficients the $A_{i}\in\mathbb{C}$, of all the elements in the null space of $M$ as
\begin{align}
Y_{\left|k\right|} & =\left(0,0,0,0,A_{1},A_{2}\right)^{T}\,.
\end{align}
Therefore, in the solution the only non-trivial values are in the
last two slots occupied by the tensor perturbations, as expected.
Also, notice that the maximum value of the subscript index of the $A_{i}$
indicates the number of PDoF of the theory.\medskip

In Appendix \ref{sec:Solutions-and-branching} we use this approach to categorically analyse each possible case that ultimately assumes a non-degenerate principal polynomial. These results are summarized in Table \ref{tab:GW_DoF}. Here, we present the various classes of sub-cases through various conditions together with a thorough classification of their PDoF by means of their SVT decomposition as well as whether each sub-case assumes a massless or massive sector or both. Case 0 revisits the situation of standard Horndeski gravity which reproduces the results in Ref.~\cite{Hou:2017bqj,Gong:2018ybk} but we further find an extra sub-case that only entails a massless sector. In this sector only the tensor perturbations propagate and in this way we exhaustively complete the standard Horndeski scenario. In case 1, we develop the general case of the full BDLS scenario where none of the contributions from the action in Eq.~\eqref{action} vanish. This case corresponds to 7 PDoF, which can be seen as two extra scalars and a vector in the massless sector compared to the standard Horndeski gravity. Case 2 then explores the possibility of the $\tilde{c}_1$ and $\tilde{c}_2$ vanishing which is a natural class of models that emerges from the study of principlal polynomial \eqref{eq:Pk} and leads to a rich structure of sub-classes with various expressions of PDoF. This situation is then further developed in cases 3, 5, and 7 where the combinations of vanishing $c_2$ and $c_3$ parameters are analyzed. In cases 4, 6, and 8, we take $\tilde{c}_2 = 0 = \tilde{c}_3$ and a similar series of sub-classes are developed.

\begin{center}
\begin{table}[H]
\begin{centering}
\setlength{\tabcolsep}{7pt} \renewcommand{\arraystretch}{1.7}{%
\begin{tabular}{|l|c|c|c|}
\hline
\multicolumn{1}{|c|}{\textbf{Theory}} & \multicolumn{1}{|c|}{\textbf{Case}} & \multicolumn{1}{|c|}{\textbf{PDoF}} & \multicolumn{1}{|c|}{\textbf{Lagrangian Density $\mathcal{L}_{i}$ $\left(S_{i}=\frac{1}{2\kappa^{2}}\int d^{4}x\,e\,\mathcal{L}_{i}\right)$}}\tabularnewline
\hline
\hline
GR or $f(T)$ & - & 2 & $\lc R$ or $f(T)$\tabularnewline
\hline
Horndeski & 0.I & 3 & Eqs.~(\ref{eq:LagrHorn1})-(\ref{eq:LagrHorn5})\tabularnewline
\hline
$G_{{\rm Tele}}$ & 1 & 7 & $G_{\text{Tele}}\left(\phi,X,T,T_{\text{ax}},T_{\text{vec}},I_{2},J_{1},J_{3},J_{5},J_{6},J_{8},J_{10}\right)$\tabularnewline
\hline
Generalized NGR & 2.I.b & 5 & $f(T,T_{{\rm ax}},T_{{\rm vec}})$\tabularnewline
\hline
Generalized teleparallel dark energy & 7 & 3 & $-A(\phi)T-\frac{1}{2}\partial_{\mu}\phi\partial^{\mu}\phi-V(\phi)$\tabularnewline
\hline
Generalized Teleparallel Scalar Tensor & 7 & 3 & $F(\phi)T+P(\phi,X)-G_{3}(\phi,X)\Box\phi$\tabularnewline
\hline
Tachyonic teleparallel gravity & 7 & 3 & $f(T,X,\phi)$\tabularnewline
\hline
\end{tabular}}
\par\end{centering}
\caption{These are some of the literature models shown against our analysis, as presented in Table. \ref{tab:GW_DoF}. }
\label{tab:dof_classes}
\end{table}
\par\end{center}

In totality, BDLS theory predicts a wide range of branches which exhaustively include all forms of SVT PDoF regarding the massless sector. These branches may also include a massive sector which is always comprised of a single (massive) scalar field, a common feature just like with the standard (scalar-tensor) Horndeski scenario. The branches of the BDLS theory are either just the massless sector or can be a combination of both the massless and massive sectors. The standard Horndeski theory, on top of the new massless branch is explored in Case 0.II, which also falls in the BDLS branch of Case 7. An interesting feature of this branching is that there are cases distinct from GR and standard Horndeski which, although enjoy the same number and type of PDoF, they are actually represent a super-class of the aforementioned standard paradigms.

\section{Polarizations of GWs}\label{sec:gw_polarizations}

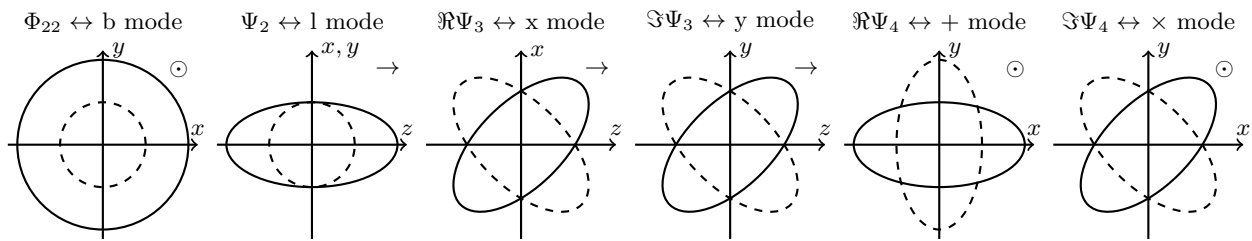
\begin{figure}[H]
\begin{tikzpicture}[thick,scale=0.25]

\foreach \x in {0,11,...,55} \draw[->] (\x,5) to +(10,0);
\foreach \x in {5,16,...,60} \draw[->] (\x,0) to +(0,10);

\node[above] at (5,10.5) {$\Phi_{22}\leftrightarrow$ b mode};
\node[above] at (10,5) {$x$};
\node[right] at (5,10) {$y$};
\node at (9,9) {$\odot$};
\draw[dashed] (5,5) circle [radius=2.25];
\draw (5,5) circle [radius=4.5];

\node[above] at (16,10.5) {$\Psi_{2}\leftrightarrow$ l mode};
\node[above] at (21,5) {$z$};
\node[right] at (16,10) {$x,y$};
\node at (20,9) {$\rightarrow$};
\draw (16,5) ellipse [x radius=4.5,y radius=2.25];
\draw[dashed] (16,5) circle [radius=2.25];

\node[above] at (27,10.5) {$\Re\Psi_{3}\leftrightarrow$ x mode};
\node[above] at (32,5) {$z$};
\node[right] at (27,10) {$x$};
\node at (31,9) {$\rightarrow$};
\draw[rotate around={45:(27,5)}] (27,5) ellipse [x radius=4.5,y radius=2.25];
\draw[dashed,rotate around={45:(27,5)}] (27,5) ellipse [x radius=2.25,y radius=4.5];

\node[above] at (38,10.5) {$\Im\Psi_{3}\leftrightarrow$ y mode};
\node[above] at (43,5) {$z$};
\node[right] at (38,10) {$y$};
\node at (42,9) {$\rightarrow$};
\draw[rotate around={45:(38,5)}] (38,5) ellipse [x radius=4.5,y radius=2.25];
\draw[dashed,rotate around={45:(38,5)}] (38,5) ellipse [x radius=2.25,y radius=4.5];

\node[above] at (49,10.5) {$\Re\Psi_{4}\leftrightarrow$ $+$ mode};
\node[above] at (54,5) {$x$};
\node[right] at (49,10) {$y$};
\node at (53,9) {$\odot$};
\draw (49,5) ellipse [x radius=4.5,y radius=2.25];
\draw[dashed] (49,5) ellipse [x radius=2.25,y radius=4.5];

\node[above] at (60,10.5) {$\Im\Psi_{4}\leftrightarrow$ $\times$ mode};
\node[above] at (65,5) {$x$};
\node[right] at (60,10) {$y$};
\node at (64,9) {$\odot$};
\draw[rotate around={45:(60,5)}] (60,5) ellipse [x radius=4.5,y radius=2.25];
\draw[dashed,rotate around={45:(60,5)}] (60,5) ellipse [x radius=2.25,y radius=4.5];

\end{tikzpicture}
\caption{All possible polarizations of GW travelling in z-direction, starting with the scalar modes breathing, longitudinal, the vector modes x,y, and the tensor modes $+,\times$. The GW deforms a sphere of freely falling test particles.}
\label{fig_polarizations}
\end{figure}

It is a well known fact that metric theories of gravity allow a maximum of six GW polarizations \cite{Eardley:1973br,Eardley:1974nw}. These six polarizations are classified according to their helicity as two tensor (helicity $\pm 2$) modes plus (+) and cross ($\times$), two vector (helicity $\pm 1$) modes called x and y and two scalar (helicity 0) modes named breathing and longitudinal modes. These polarizations are further illustrated in Fig.~\ref{fig_polarizations}. We can probe the polarization content of GWs by measuring the outputs of their relative amplitudes in the detectors \cite{Will:2014kxa,Chatziioannou:2012rf,Isi:2017equ,Callister:2017ocg,Isi:2017fbj}.\medskip

At the moment there is a three-detector network that can be used to distinguish among very specific subsets of all possible polarization combinations. On the other hand, to more accurately reconstruct the polarization content of a GW we would need a five-detector network also determine the effect of degeneracies. Thus, future measurements will shed more light on the status on the types of polarization content. Using a simplified analysis that strictly relies on GR templates \cite{Isi:2017fbj}, there have been reported some constraints on pure combinations such as only tensor against full vector and full tensor against full scalar in \cite{Abbott:2017oio,Abbott:2018lct,LIGOScientific:2019fpa}. It turned out that in this analysis using (GW170814, GW170817, and GW170818) only the tensorial polarizations were not disfavoured.\medskip

Let us point out that this analysis states that a GW cannot only have scalar or only vector polarizations but is highly more likely to have only tensor polarizations. Still, it does not exclude theories that can have a combination of tensor with scalar--vector polarizations. It also does not exclude GWs produced with only vector or only scalar polarizations directly from their source. Thus, a viable theory must predict at the very least tensor polarizations.\medskip

The way we measure the polarizations is directly linked to the electric components of the linearized Riemann tensor $\lc{R}_{i0j0}$. These components control the response of test particles in the presence of a GW. In analytical terms this is described through the geodesic deviation equation Ref.~\cite{Carroll:2004st}
\begin{equation}
    \ddot{x}_i = -\lc{R}_{i0j0} x^j\,,\label{eq:geodesic_dev}
\end{equation}
where dots represent coordinate time derivatives, $(t,x,y,z) = (0,1,2,3)$, $i = \lbrace 1,2,3 \rbrace$ and $x^j = (x, y, z)$. As a matter of fact there is a very useful tool in studying the polarizations of massless (and almost massless) GWs which is based on the Newman-Penrose (NP) formalism called the E(2) classification. This tool facilitates the categorization of the allowed polarizations of massless gravitational waves in a given gravitational theory via the representation of the little group, which is the two-dimensional Euclidean group $\mathrm{E}(2)$ \cite{Eardley:1973br,Eardley:1974nw,Will:1993ns}. Thus, within the E(2) framework we can parametrize the six independent components of $\lc{R}_{i0j0}$, using the NP variables as
\begin{align}
    \lc{R}{}_{0i0j} & =\left(\arraycolsep=2pt\def\arraystretch{1.5}\begin{array}{ccc}
    \frac{1}{2}(\Re\Psi_{4}+\Phi_{22}) & \frac{1}{2}\Im\Psi_{4} & -2\Re\Psi_{3}\\
    \frac{1}{2}\Im\Psi_{4} & -\frac{1}{2}(\Re\Psi_{4}-\Phi_{22}) & 2\Im\Psi_{3}\\
    -2\Re\Psi_{3} & 2\Im\Psi_{3} & -6\Psi_{2}
    \end{array}\right)\,.\label{eq:RiemannELECTRIC_E2}
\end{align}
where $\Phi_{22},\Psi_{2},\Psi_{3},\Psi_{4}$ are the NP variables,  $\Re$ represents the real part and $\Im$ the imaginary one. These NP variables can also be classified with respect to their helicity states through
\begin{equation}
\arraycolsep=2pt\def\arraystretch{1.5}\begin{array}{lllllc}
\Psi_{2} & : & s=0\,, & \Phi_{22} & : & s=0\,,\\
\Psi_{3} & : & s=-1\,, & \overline{\Psi}_{3} & : & s=1\,,\\
\Psi_{4} & : & s=-2\,, & \overline{\Psi}_{4} & : & s=2\,,
\end{array}\label{eq:POL_AMPL_helicities}
\end{equation}
where the overbar denotes complex conjugation. One can directly deduce from
Eq.~(\ref{eq:POL_AMPL_helicities}) that $\Phi_{22},\Psi_{2}$
are related to scalar DoF, $\Psi_{3}$ is related to vectorial DoF
and finally $\Psi_{4}$ is related to tensorial DoF. Finally, we can visualize the parametrization of Eq.~(\ref{eq:RiemannELECTRIC_E2}) as in Fig.~\ref{fig_polarizations}.
\begin{center}
\begin{table}[ht]
\begin{centering}
\setlength{\tabcolsep}{5.5pt} \renewcommand{\arraystretch}{1.5}{%
\begin{tabular}{|c|l|c|c|c|c|c|c||c|c|}
\hline
\multirow{5}{*}{\textbf{Cases}} & \multicolumn{1}{c|}{\multirow{5}{*}{\textbf{Conditions}}} & \multicolumn{8}{c|}{\textbf{Polarizations}}\tabularnewline
\cline{3-10} \cline{4-10} \cline{5-10} \cline{6-10} \cline{7-10} \cline{8-10} \cline{9-10} \cline{10-10}
 &  & \multicolumn{6}{c||}{Massless sector} & \multicolumn{2}{c|}{Massive sector}\tabularnewline
 &  & \multicolumn{6}{c||}{$\omega^{2}=k{}^{2}$} & \multicolumn{2}{c|}{$\omega^{2}-k{}^{2}=m^{2}$}\tabularnewline
\cline{3-10} \cline{4-10} \cline{5-10} \cline{6-10} \cline{7-10} \cline{8-10} \cline{9-10} \cline{10-10}
 &  & \multicolumn{2}{c|}{Scalar} & \multicolumn{2}{c|}{Vector} & \multicolumn{2}{c||}{Tensor} & \multicolumn{2}{c|}{Scalar}\tabularnewline
\cline{3-10} \cline{4-10} \cline{5-10} \cline{6-10} \cline{7-10} \cline{8-10} \cline{9-10} \cline{10-10}
 &  & b & l & x & y & $+$ & $\times$ & b & l\tabularnewline
\hline
\multirow{1}{*}{-} & \multirow{1}{*}{$G_{4}-G_{\text{Tele},T}\neq0$} & \multirow{1}{*}{-} & \multirow{1}{*}{-} & \multirow{1}{*}{-} & \multirow{1}{*}{-} & \multirow{1}{*}{$\surd$} & \multirow{1}{*}{$\surd$} & \multirow{1}{*}{-} & \multirow{1}{*}{-}\tabularnewline
\hline
\multirow{2}{*}{0} & $G_{\text{Tele},T_{\text{vec}}}=0,\,G_{\text{Tele},T_{\text{ax}}}=0,\, G_{\text{Tele},I_{2}}=0\,,$ & \multicolumn{6}{c||}{} & \multicolumn{1}{c}{} & \tabularnewline
 &
$ G_{\text{Tele},X}=0,G_{\text{Tele},\phi\phi}=0$& \multicolumn{6}{c||}{} & \multicolumn{1}{c}{} & \tabularnewline
\hline
\multirow{2}{*}{0.I} & $G_{2,\phi\phi}\neq0$ and  & \multirow{2}{*}{-} & \multirow{2}{*}{-} & \multirow{2}{*}{-} & \multirow{2}{*}{-} & \multirow{2}{*}{$\surd$} & \multirow{2}{*}{$\surd$} & \multirow{2}{*}{$\surd$} & \multirow{2}{*}{$\surd$}\tabularnewline
 &
 $-3G_{4,\phi}{}^{2}+(G_{4}-G_{\text{Tele},T})\left(2G_{3,\phi}-G_{2,X}\right)\neq0$&  &  &  &  &  &  &  & \tabularnewline
\hline
0.II & $G_{2,\phi\phi}=0$ and $-3G_{4,\phi}{}^{2}+(G_{4}-G_{\text{Tele},T})\left(2G_{3,\phi}-G_{2,X}\right)=0$ & - & - & - & - & $\surd$ & $\surd$ & - & -\tabularnewline
\hline
1 & $G_{\text{Tele},T_{\text{vec}}}\neq0,\,G_{\text{Tele},T_{\text{ax}}}\neq0,\,\tilde{c}_{1}\neq0,\,\tilde{c}_{2}\neq0$ & $\surd$ & - & - & - & $\surd$ & $\surd$ & $\surd$ & $\surd$\tabularnewline
\hline
2 & $G_{\text{Tele},T_{\text{vec}}}\neq0,\,G_{\text{Tele},T_{\text{ax}}}\neq0,\,\tilde{c}_{1}=0,\,\tilde{c}_{2}=0$ & \multicolumn{1}{c}{} & \multicolumn{1}{c}{} & \multicolumn{1}{c}{} & \multicolumn{1}{c}{} & \multicolumn{1}{c}{} &  & \multicolumn{1}{c}{} & \tabularnewline
\cline{1-2} \cline{2-2}
2.I & $G_{\text{Tele},T_{\text{ax}}}\neq0,\tilde{c}_{3}\neq0,\tilde{c}_{4}\neq0$ & \multicolumn{1}{c}{} & \multicolumn{1}{c}{} & \multicolumn{1}{c}{} & \multicolumn{1}{c}{} & \multicolumn{1}{c}{} &  & \multicolumn{1}{c}{} & \tabularnewline
\hline
2.I.a & $G_{\text{Tele},T_{\text{ax}}}\neq0,\,\tilde{c}_{3}\neq0,\,\tilde{c}_{4}\neq0$ & - & - & - & - & $\surd$ & $\surd$ & - & -\tabularnewline
\hline
2.I.b & $ c\neq0,\,\tilde{c}_{3}=0,\,\tilde{c}_{4}=0 $ & - & - & - & - & $\surd$ & $\surd$ & - & -\tabularnewline
\hline
2.II & $G_{3,\phi}=Z_2,\,G_{2,\phi\phi}=-G_{\text{Tele},\phi\phi}$ & \multicolumn{1}{c}{} & \multicolumn{1}{c}{} & \multicolumn{1}{c}{} & \multicolumn{1}{c}{} & \multicolumn{1}{c}{} &  & \multicolumn{1}{c}{} & \tabularnewline
\hline
2.II.a & $G_{\text{Tele},I_{2}}-2G_{4,\phi}\neq0$ & $\surd$ & - & - & - & $\surd$ & $\surd$ & - & -\tabularnewline
\hline
2.II.b & $G_{\text{Tele},I_{2}}-2G_{4,\phi}=0$ & - & - & - & - & $\surd$ & $\surd$ & - & -\tabularnewline
\hline
3 & $G_{\text{Tele},T_{\text{vec}}}\neq0,\,G_{\text{Tele},T_{\text{ax}}}=0,\,\tilde{c}_{1}\neq0,\,\tilde{c}_{2}\neq0$ & $\surd$ & - & - & - & $\surd$ & $\surd$ & $\surd$ & $\surd$\tabularnewline
\hline
4 & $G_{\text{Tele},T_{\text{vec}}}\neq0,\,G_{\text{Tele},T_{\text{ax}}}=0,\,\tilde{c}_{1}=0,\,\tilde{c}_{2}=0$ & \multicolumn{1}{c}{} & \multicolumn{1}{c}{} & \multicolumn{1}{c}{} & \multicolumn{1}{c}{} & \multicolumn{1}{c}{} &  & \multicolumn{1}{c}{} & \tabularnewline
\cline{1-2} \cline{2-2}
4.I & $G_{\text{Tele},T_{\text{vec}}}=-\tfrac{2}{3}(G_{4}-G_{\text{Tele},T}),\,G_{4,\phi}=\tfrac{1}{2}G_{\text{Tele},I_{2}}$ & \multicolumn{1}{c}{} & \multicolumn{1}{c}{} & \multicolumn{1}{c}{} & \multicolumn{1}{c}{} & \multicolumn{1}{c}{} &  & \multicolumn{1}{c}{} & \tabularnewline
\hline
4.I.a & $\tilde{c}_{3}\neq0,\tilde{c}_{4}\neq0$ & - & - & - & - & $\surd$ & $\surd$ & - & -\tabularnewline
\hline
4.I.b & $\tilde{c}_{3}=0,\tilde{c}_{4}=0$ & - & - & - & - & $\surd$ & $\surd$ & - & -\tabularnewline
\hline
4.II & $G_{3,\phi}=Z_2,\, G_{2,\phi\phi}=-G_{\text{Tele},\phi\phi}$ & \multicolumn{1}{c}{} & \multicolumn{1}{c}{} & \multicolumn{1}{c}{} & \multicolumn{1}{c}{} & \multicolumn{1}{c}{} &  & \multicolumn{1}{c}{} & \tabularnewline
\hline
4.II.a & $G_{\text{Tele},I_{2}}-2G_{4,\phi}\neq0$ & $\surd$ & - & - & - & $\surd$ & $\surd$ & - & -\tabularnewline
\hline
4.II.b & $G_{\text{Tele},I_{2}}-2G_{4,\phi}=0$ & - & - & - & - & $\surd$ & $\surd$ & - & -\tabularnewline
\hline
5 & $G_{\text{Tele},T_{\text{vec}}}=0,\,G_{\text{Tele},T_{\text{ax}}}\neq0,\,\tilde{c}_{1}\neq0,\,\tilde{c}_{2}\neq0$ & - & - & - & - & $\surd$ & $\surd$ & $\surd$ & $\surd$\tabularnewline
\hline
6 & $G_{\text{Tele},T_{\text{vec}}}=0,\,G_{\text{Tele},T_{\text{ax}}}\neq0,\,\tilde{c}_{1}=0,\,\tilde{c}_{2}=0$ & - & - & - & - & $\surd$ & $\surd$ & - & -\tabularnewline
\hline
7 & $G_{\text{Tele},T_{\text{vec}}}=0,\,G_{\text{Tele},T_{\text{ax}}}=0,\,\tilde{c}_{1}\neq0,\,\tilde{c}_{2}\neq0$ & - & - & - & - & $\surd$ & $\surd$ & $\surd$ & $\surd$\tabularnewline
\hline
8 & $G_{\text{Tele},T_{\text{vec}}}=0,\,G_{\text{Tele},T_{\text{ax}}}=0,\,\tilde{c}_{1}=0,\,\tilde{c}_{2}=0$ & - & - & - & - & $\surd$ & $\surd$ & - & -\tabularnewline
\hline
\end{tabular}}
\par\end{centering}
\caption{All Branches of the BDLS theory with corresponding Polarizations. The quantity $Z_2$ is defined in the appendix (see Eqs.~\eqref{Z1}-\eqref{Z2}) and the quantities $\tilde{c}_{i}$ are defined in~\eqref{eq:c1tdef},~\eqref{eq:c2tdef},~\eqref{eq:c3tdef} and \eqref{eq:c4tdef}, respectively.}
\label{tab:GW_POL}
\end{table}
\par\end{center}

For our analysis, we will use the SVT decomposition of the electric components of the Riemann tensor Eq.~(\ref{eq:geodesic_dev}) which in terms of gauge invariant variables (see Eq.~\eqref{eq:gauge_transformations}) assumes the form
\begingroup
\begin{align}
\lc{R}{}_{0i0j} & =\left(
\arraycolsep=2pt\def\arraystretch{1.5}\begin{array}{ccc}
\ddot{\psi}-\tfrac{1}{2}\ddot{h}_{+} & -\tfrac{1}{2}\ddot{h}_{\times} & -\tfrac{1}{2}ik(\dot{\beta}_{1}+\dot{\Lambda}_{1})\\
-\tfrac{1}{2}\ddot{h}_{\times} & \ddot{\psi}+\tfrac{1}{2}\ddot{h}_{+} & -\tfrac{1}{2}ik(\dot{\beta}_{2}+\dot{\Lambda}_{2})\\
-\tfrac{1}{2}ik(\dot{\beta}_{1}+\dot{\Lambda}_{1}) & -\tfrac{1}{2}ik(\dot{\beta}_{2}+\dot{\Lambda}_{2}) & \ddot{\psi}-k^{2}(\dot{\chi}+\Phi)
\end{array}\right)\,.\label{eq:ERiemSVT}
\end{align}
\endgroup

In contrast to the NP representation in Eq.~(\ref{eq:RiemannELECTRIC_E2}), the SVT counterpart Eq.~(\ref{eq:ERiemSVT}) is valid for both a massless and massive GWs. As a matter of fact when the GW is massless then Eq.~(\ref{eq:ERiemSVT}) and Eq.~(\ref{eq:RiemannELECTRIC_E2}) will coincide.\medskip

Following the analysis presented in Sec.~\ref{sec:dof_analysis} and also using the electric components of the Riemann tensor as formulated in Eq.~(\ref{eq:ERiemSVT}), we exhaustively calculated all possible polarizations contained in the BDLS theory. The calculation process, at this stage, is straightforward since we only had to substitute the solutions $Y_{\left|k\right|}$ and $Y_{\left|m\right|}$, as already illustrated in Appendix \ref{sec:Solutions-and-branching}, into Eq.~(\ref{eq:ERiemSVT}). Then, we gathered all the results into Table \ref{tab:GW_POL} which closely follows the structure of the PDoF Table \ref{tab:GW_DoF}.\medskip

We first considered the most trivial cases of GR and $f(T)$ gravity theories where we properly reproduced the well known and only tensor polarizations. Following up with the standard Horndeski Case 0, for the first branch Case 0.I, which assumes 3 PDoF, we found tensor polarizations for the massless sector and a mix of breathing and longitudinal modes for the massive sector. Regarding the new branch labeled Case 0.II, we found only tensor polarizations, as expected, since only tensorial PDoF were found in Sec.~\ref{sec:dof_analysis}.\medskip

Delving into the TG realm, in Case 1 (the full BDLS theory), we found the breathing mode along the usual tensor modes in the massless section and the breathing along with the longitudinal for the massive sector. In terms of the polarizations content the full BDLS theory has an extra breathing mode for the massless sector compared to the standard full Horndeski (Case 0.I). This unique polarization imprint is only shared between Case 1 and Case 3 inside the BDLS class although Case 1 includes one additional massless scalar and one massless vectorial PDoF.\medskip

Another interesting type of polarization imprint was produced in Cases 2.I.a, 2.I.b, 2.II.b, 5 and 6 where although there exists a massless scalar PDoF it does not leave a polarization imprint. This also happens for Cases 2.I.a and 4.I.a for a massive scalar DoF. The elusiveness of DoF also happens for the vectorial DoFs found in Cases 1, 2.I.a, 2.I.b, 2.II.a and 2.II.b, which are the only cases predicting vectorial DoF. This, phenomenon of a DoF not leaving a polarization imprint is directly linked to the fact that it is not coupled to the corresponding metrical DoF to which the Riemann tensor is sensitive to.\medskip

We also have a few trivial subclasses described by Cases 2.I.b, 2.II.b, 4.I.b, 4.II.b, 6 and 8 where only tensor polarizations were found. This, is also not so straightforward to expect since except for the Cases 4.I.b, 4.II.b and 8 which only enjoy tensorial PDoF, the cases 2.I.b, 2.II.b and 6 predict more than 2 DoF as as explained in Sec.~\ref{sec:dof_analysis}.\medskip

On the whole, in the BDLS theory although scalar, vectorial and tensorial propagating DoF are predicted as illustrated in Sec.~\ref{sec:dof_analysis} we can have only combinations of scalar, tensor polarizations for the massless sectors and only scalar ones for the massive sectors. Thus, although vectorial propagating DoF are predicted they leave no polarization imprint.\medskip

\section{Conclusion}\label{sec:conclusion}

This work explores the number and nature of the PDoF associated with BDLS theory in a Minkowski background. In this context, we also expose the expressions of GW polarization through the various subclasses of models of the model. BDLS theory is a TG analogous formulation of Horndeski gravity in which the connection is changed from Levi-Civita, which is curvature-based, to the teleparallel connection, which is torsional. This simple change drastically impacts the breath of the theory by adding a term in the Lagrangian \eqref{action} which starkly increases the potential models that ensue. We cover this background in Sec.~\ref{sec:intro_BDLS} where we also introduce the gravitational tools in which BDLS theory is constructed.\medskip

In Sec.~\ref{sec:perturbations}, we then describe the transition to tetrad based perturbations. The use of metric tensor perturbations is well known \cite{Mukhanov:991646}. However, given that metrics do not define unique tetrads, we must be careful to define tetrads that not only produce the metric but also that continue to satisfy the field equations (up to perturbative order). In Eqs.~(\ref{eq:tetrad_pert},\ref{eq:phi_pert}) we prescribe the tetrad together with its scalar field perturbation which reproduces the correct perturbed metric in Eq.~(\ref{eq:metric_pert}). The equations of motion are then determined (\ref{Wmunutotal},\ref{Wscalartotal}) and simplified into an appropriate form using symmetry operators (\ref{eq:FE_symm},\ref{eq:FE_antisymm}).\medskip

The perturbed field equations are explored in Sec.~\ref{sec:dof_analysis} where the tetrad is decomposed into its SVT constituents \eqref{eq:tetrad_perturbation} which can be correlated with their the usual SVT decomposed metric \eqref{eq:delta_metric_perturbation}. Through this prism, we explore the number of PDoF which is an important property to determine to accurately understand how this freedom is then expressed. In Table \ref{tab:GW_DoF} we display the nature and number of DoF and determine whether they are massive or massless for the various subclasses of models within BDLS theory (which expands on the Horndeski branches). The details of these tedious calculations are shown in Appendix \ref{sec:Solutions-and-branching} where each branch is explained.\medskip

It is important to highlight that we identify a number of cases in which extra PDoF are identified where we find a maximum of seven PDoF for one of these cases. In general, we find both massive and massless cases and a distribution over the scalar--vector--tensor modes for these DoF. In the massive instance, we naturally find that this is associated with the scalar field. To better represent the impact of these results, we list how this will effect a number of well-known TG theories that are second order in Table~\ref{tab:dof_classes}. While some of these have already been determined elsewhere, we are in agreement with literature results. The prism of the BDLS derivation presented here is that it can also collect these results together in one formalism.\medskip

PDoF can be expressed as GW polarization which we explore in Sec.~\ref{sec:gw_polarizations}. Here, we describe the possible SVT polarizations of GWs as well as the way in which they can be determined though the Levi-Civita Riemann tensor quantity in the geodesic deviation equation \eqref{eq:geodesic_dev}. The results are presented in Table \ref{tab:GW_DoF} where we describe the polarization characteristics for each of the subclasses defined in Table~\ref{tab:GW_POL}. Of particular interest, we expand on the work done in standard Horndeski gravity in which the massive branch was explored in Ref.~\cite{Hou:2017bqj} but where the massless case is omitted, which we expand upon here.\medskip

We also find that a maximum of four polarization modes can be produced by BDLS theory and that none of them express a vector mode. Despite this, we continue to have massless and massive scalar mode cases. Beyond the standard Horndeski gravity, we find a wide variety on polarization states in these subclasses which may be interesting for future polarization detection analyses in the coming years.\medskip

GW polarization is a key test of gravity that is yet to be exposed to the same level of details as other features of gravity such as the speed of propagation of GW radiation. In general, TG has not been exhaustively studied but has shown promise in a number of interesting areas of astrophyscs \cite{Bamba:2012cp,Linder:2010py,Escamilla-Rivera:2019ulu,Finch:2018gkh,Briffa:2020qli,LeviSaid:2020mbb,Farrugia:2016xcw,Farrugia:2016qqe,Bahamonde:2020bbc}. As expressed in this work, BDLS theory offers a wider landscape on which to produce second-order theories of gravity where only one scalar field is incorporated. It would be interesting to further study the features of this theory.

\subsection*{Acknowledgements}
J.L.S. would also like to acknowledge funding support from Cosmology@MALTA which is supported by the University of Malta.  S.B. and M.H. were supported by the Estonian Research Council grants PRG356 ``Gauge Gravity"  and by the European Regional Development Fund through the Center of Excellence TK133 ``The Dark Side of the Universe". The authors would like to acknowledge networking support by the COST Action CA18108. The research in this work is partially funded by the Tertiary Education Scholarship Scheme (TESS, Malta).

\appendix

\section{Expansion of Multivariable Function}
\label{Expansion_Multivariable_Function}

The perturbative expansion of the field equations is obtained by first expanding the functions in the Lagrangian. This is done by performing a Taylor expansion about background values. Consider a function of scalars, say $G(\alpha,\beta)$, such that the parameters of the function are expanded as follows
\begin{align}
    \alpha &= \alpha^{(0)} + \epsilon\, \alpha^{(1)} + \epsilon^{2} \alpha^{(2)}\,,\\
    \beta &= \beta^{(0)} + \epsilon\, \beta^{(1)} + \epsilon^{2} \beta^{(2)} + \epsilon^{3} \beta^{(3)}\,,
\end{align}
where $\alpha^{(0)}$ and $\beta^{(0)}$ are constants at the zeroth order which represent the background parameters. Taylor expanding function $G$ about the zeroth order results in the following equation:
\begin{align}
    G(\alpha,\beta) &= G(0) + G_{,\alpha}(0) \left[\alpha - \alpha^{(0)}\right] + G_{,\beta}(0) \left[\beta - \beta^{(0)}\right] + G_{,\alpha\beta}(0)\left[\alpha - \alpha^{(0)}\right]\left[\beta - \beta^{(0)}\right]\\ \nonumber & \quad + \frac{1}{2} G_{,\alpha\alpha}(0) \left[\alpha - \alpha^{(0)}\right]^{2} + \frac{1}{2} G_{,\beta\beta}(0)\left[\beta - \beta^{(0)}\right]^{2}  + ...\\ \nonumber &= G(0) + \epsilon \left[G_{,\alpha}(0) \alpha^{(1)} + G_{,\beta}(0) \beta^{(1)}\right]\\ \nonumber & \quad + \epsilon^{2} \left[G_{,\alpha}(0) \alpha^{(2)} + G_{,\beta}(0) \beta^{(2)} + G_{,\alpha\beta}(0) \alpha^{(1)} \beta^{(1)} + \frac{1}{2} \left(G_{,\alpha\alpha}(0) {\alpha^{(1)}}^{2} + G_{,\beta\beta}(0) {\beta^{(1)}}^{2}\right)\right] + \mathcal{O}(\epsilon^{3})\,,
\end{align}
where $G(0) = G\left(\alpha^{(0)},\beta^{(0)}\right)$ stands for the background of the function $G$. This equation shows that scalars with contributions to perturbation orders higher two, such as $\beta^{(3)}$, will not appear in the second order expansion of the function. In the case of BDLS theory, the scalar invariants $J_{1}$, $J_{3}$, $J_{5}$, $J_{6}$, $J_{8}$ and $J_{10}$ do not appear in the expansion as seen in Eq.~(\ref{function_expansion}) since they do not have contributions that are smaller than the perturbative order of 3. Moreover, in the case of $X$, there is only a second order contribution which leads to no second order derivative terms in the function expansion.

\section{Solutions and Branching\label{sec:Solutions-and-branching}}

The general solution of the system in Eq.~(\ref{eq:TotalMat}) will be a linear combination of the elements of the null space of matrix $M$. We will denote the coefficients of the linear combinations as $A_{i}\in\mathbb{C}$ for the massless dispersion relation $\omega^{2}-k^{2}=0$ and $B_{i}\in\mathbb{C}$ for the massive dispersion relation $\omega^{2}-k^{2}=m^{2}$. These coefficients also act as labels for the number of PDoF. Hence each $A_{i}$ or $B_{i}$ denotes one PDoF. If there are both massless and massive branches we have a solution space for each sector and thus the sum of their dimensions is the total number PDoF. We will denote the solutions with respect to to the massless dispersion relation $\omega^{2}-k^{2}=0$ as $Y_{\left|k\right|}$ and the ones
corresponding to a massive dispersion relation $\omega^{2}-k^{2}=m^{2}$ as $Y_{\left|m\right|}$. Finally these solutions will be equal and compared with the column vector $Y$ defined in Eq.~(\ref{eq:TotalMat}).

After each solution $Y_{\left|k\right|}$ or $Y_{\left|m\right|}$ we will calculate the electric components of the Riemann tensor that correspond to each of them as $\lc{R}_{0i0j}(Y_{\left|k\right|})$ or $\lc{R}_{0i0j}(Y_{\left|m\right|})$ accordingly. Finally, we will comment only on the first few interesting cases since, although the list is exhaustive, the rest of the cases have similar or identical interpretation. \\

Let us now explore all the possible cases.\\ \\
\textbf{\underline{Case 0:} Horndeski} $\left(G_{\text{Tele},T_{\text{vec}}}=0,\,G_{\text{Tele},T_{\text{ax}}}=0,\,G_{\text{Tele},I_{2}}=0,\,G_{\text{Tele},X}=0,\,G_{\text{Tele},\phi\phi}=0\right)$\\ \\
The Horndeski case has the following determinant
\small{\begin{align}
    P(k) & =-16(G_{4}-G_{\text{Tele},T}){}^{5}k{}^{8}\bigl(\omega^{2}-k{}^{2}\bigr)^{2}\Bigl((G_{4}-G_{\text{Tele},T})G_{2,\phi\phi}+(-3G_{4,\phi}{}^{2}+(G_{4}-G_{\text{Tele},T})(2G_{3,\phi}-G_{2,X}))\bigl(\omega^{2}-k{}^{2}\bigr)\Bigr)\,,
\end{align}}
\normalsize
from which we further have two sub-cases:\\ \\
\textbf{\underline{Case 0.I}} $(G_{2,\phi\phi}\neq0$ and $-3G_{4,\phi}{}^{2}+(G_{4}-G_{\text{Tele},T})\left(2G_{3,\phi}-G_{2,X}\right)\neq0)$.\label{Case-0ii.}\\ \\
In this case the determinant reads
\small{\begin{align}
    P(k) & =-16(G_{4}-G_{\text{Tele},T}){}^{5}k{}^{8}\bigl(\omega^{2}-k{}^{2}\bigr)^{2}\Bigl((G_{4}-G_{\text{Tele},T})G_{2,\phi\phi}+(-3G_{4,\phi}{}^{2}+(G_{4}-G_{\text{Tele},T})(2G_{3,\phi}-G_{2,X}))\bigl(\omega^{2}-k{}^{2}\bigr)\Bigr)\,,
\end{align}}\normalsize
from which it is evident that it is non-degenerate and there are two
dispersion relations $\omega^{2}-k{}^{2}=0$ massless speed of light propagation and
$$(G_{4}-G_{\text{Tele},T})G_{2,\phi\phi}+(-3G_{4,\phi}{}^{2}+(G_{4}-G_{\text{Tele},T})(2G_{3,\phi}-G_{2,X}))\bigl(\omega^{2}-k{}^{2}\bigr)=0$$
which defines an effective mass as $$m^{2}=-\frac{(G_{4}-G_{\text{Tele},T})G_{2,\phi\phi}}{-3G_{4,\phi}{}^{2}+(G_{4}-G_{\text{Tele},T})(2G_{3,\phi}-G_{2,X}))}>0\,,$$ and then
\begin{align}
    Y_{\omega} & =\left(\delta\phi,\psi,\Phi,\beta{}_{i},h{}_{ij}\right)\,,\label{eq:case0ifields}\\\nonumber \\
    Y_{\left|k\right|} & =\left(0,0,0,0,0,\frac{2A_{1}}{k^{2}},\frac{2A_{2}}{k^{2}}\right)^{T}\,,\label{eq:case0imassless}\\\nonumber \\
    \lc{R}_{0i0j}(Y_{\left|k\right|}) & =\left(\begin{array}{ccc}
    A_{1} & A_{2} & 0\\
    A_{2} & -A_{1} & 0\\
    0 & 0 & 0
    \end{array}\right)\,,\label{eq:ERiemcase0imassless}\\\nonumber \\
    Y_{\left|m\right|} & =\left(2(G_{4}-G_{\text{Tele},T})B_{1},-G_{4,\phi}B_{1},G_{4,\phi}B_{1},0,0,0,0\right)^{T},\label{eq:case0imassive}\\\nonumber \\
    \lc{R}_{0i0j}(Y_{\left|m\right|}) & =G_{4,\phi}\left(\begin{array}{ccc}
    \bigl(m^{2}+k{}^{2}\bigr)B_{1} & 0 & 0\\
    0 & \bigl(m^{2}+k{}^{2}\bigr)B_{1} & 0\\
    0 & 0 & m^{2}B_{1}
    \end{array}\right)\,.\label{eq:ERiemcase0imassive}
\end{align}
The full Horndeski theory assumes 3 PDoF, 2 of which are the tensor modes described by the parameters $\left(A_{1},A_{2}\right)$ in the massless sector Eq.~(\ref{eq:case0imassless}) and the remaining DoF is a scalar described by the parameter $B_{1}$ in the massive sector Eq.~(\ref{eq:case0imassive}). It is evident that by setting $G_{4,\phi}=0$ one can completely hide the massive scalar from the polarization detectors Eq.~(\ref{eq:ERiemcase0imassive}) but it will still propagate as can be seen from Eq.~(\ref{eq:case0imassive}). \\ \\
\textbf{\underline{Case 0.II}} ($G_{2,\phi\phi}=0$ and $-3G_{4,\phi}{}^{2}+(G_{4}-G_{\text{Tele},T})\left(2G_{3,\phi}-G_{2,X}\right)=0$).\\ \\
A solution of this system
that covers the whole solution manifold is $$G_{2,X}=-\frac{3G_{4,\phi}{}^{2}}{(G_{4}-G_{\text{Tele},T})}+2G_{3,\phi}\,,$$ and then the determinant becomes
\begin{align}
    P(k) & =-8(G_{4}-G_{\text{Tele},T}){}^{5}G_{4,\phi}k{}^{8}\bigl(\omega^{2}-k{}^{2}\bigr)^{2}\,,
\end{align}
and for $G_{4,\phi}\neq0$ the solution is
\begin{align}
    Y_{\omega} & =\left(\delta\phi,\psi,\beta{}_{i},h{}_{ij}\right)\,,\\\nonumber \\
    Y_{\left|k\right|} & =\left(0,0,0,0,\frac{2A_{1}}{k^{2}},\frac{2A_{2}}{k^{2}}\right)^{T}\,,\label{eq:case0iimassless1}
\end{align}
whereas for $G_{4,\phi}=0$
\begin{align}
    Y_{\omega} & =\left(\psi,\beta{}_{i},h{}_{ij}\right)\,,\\ \nonumber \\
    Y_{\left|k\right|} & =\left(0,0,0,\frac{2A_{1}}{k^{2}},\frac{2A_{2}}{k^{2}}\right)^{T},\label{eq:case0iimassless2}
\end{align}
and thus for both cases $G_{4,\phi}=0$ and $G_{4,\phi}\neq0$
\begin{align}
    \lc{R}_{0i0j}(Y_{\left|k\right|}) & =\left(\begin{array}{ccc}
    A_{1} & A_{2} & 0\\
    A_{2} & -A_{1} & 0\\
    0 & 0 & 0
    \end{array}\right),\label{eq:case0iimasslessERiem}
\end{align}
Although the algebraic solutions are different as can be seen from Eq.~(\ref{eq:case0iimassless1}) and Eq.~(\ref{eq:case0iimassless2}) the final physical solution is the same, i.e only tensor modes. This is also reflected in the polarizations described by Eq.~(\ref{eq:case0iimasslessERiem}) since their value does not depend at all by the choice of $G_{4,\phi}$.\\ \\
\textbf{\underline{Case 1: Full BDLS theory} }$ (G_{\text{Tele},T_{\text{vec}}}\neq0,\,G_{\text{Tele},T_{\text{ax}}}\neq0,\,\tilde{c}_{1}\neq0,\,\tilde{c}_{2}\neq0)$\\ \\
In this case
 the determinant reads
\begin{align}
    P(k) & =-\frac{16384}{243}(G_{4}-G_{\text{Tele},T}){}^{5}G_{\text{Tele},T_{\text{vec}}}{}^{3}G_{\text{Tele},T_{\text{ax}}}{}^{3}k{}^{12}\bigl(\omega^{2}-k{}^{2}\bigr)^{8}\Bigl(\tilde{c}_{1}+\tilde{c}_{2}\bigl(\omega^{2}-k{}^{2}\bigr)\Bigr),
\end{align}
where $\tilde{c}_{1}$ and $\tilde{c}_{2}$ are defined in Eq.~(\ref{eq:c1tdef})-Eq.~(\ref{eq:c2tdef}). It is evident that it is non-degenerate and there are two dispersion relations $\omega^{2}-k{}^{2}=0$ which describes massless speed 1 propagation and $\tilde{c}_{1}+\tilde{c}_{2}\bigl(\omega^{2}-k{}^{2}\bigr)=0$ which defines an effective mass as $m^{2}=-\frac{\tilde{c}_{1}}{\tilde{c}_{2}}>0$. The solutions for this case are
\begin{align}
    Y_{\omega} & =\left(\delta\phi,\psi,\Phi,\chi,\sigma,\beta{}_{j},\Sigma{}_{j},\Lambda{}_{j},h{}_{ij}\right)\,,\\\nonumber \\
    Y_{\left|k\right|} & =\Big(0,-\frac{A_{1}}{k^{2}},-\frac{2A_{1}}{G_{\text{Tele},T_{\text{vec}}}k{}^{2}}(G_{4}-G_{\text{Tele},T}+G_{\text{Tele},T_{\text{vec}}}),\nonumber\\
    &\,\,\,\frac{iA_{1}}{G_{\text{Tele},T_{\text{vec}}}\left|k\right|{}^{3}}(2(G_{4}-G_{\text{Tele},T})+3G_{\text{Tele},T_{\text{vec}}}),A_{2},-A_{3},-A_{4},i\left|k\right|A_{3},i\left|k\right|A_{4},A_{3},A_{4},\frac{2A_{5}}{k^{2}},\frac{2A_{6}}{k^{2}}\Big)^{T}\,,\label{eq:case1massless}\\\nonumber \\
    \lc{R}_{0i0j}(Y_{\left|k\right|}) & =\left(\begin{array}{ccc}
    A_{1}+A_{5} & A_{6} & 0\\
    A_{6} & A_{1}-A_{5} & 0\\
    0 & 0 & 0
    \end{array}\right)\,,\label{eq:ERiemcase1massless}\\\nonumber \\
    Y_{\left|m\right|} & =\Big(-2(2(G_{4}-G_{\text{Tele},T})+3G_{\text{Tele},T_{\text{vec}}})B_{1},-(G_{\text{Tele},I_{2}}-2G_{4,\phi})B_{1},(G_{\text{Tele},I_{2}}-2G_{4,\phi})B_{1},\nonumber\\
    &\,\,\,\,\,\,\,\, \,\,\,\,\,\,0,0,0,0,0,0,0,0,0,0\Big)^{T}\,,\label{eq:case1massive}\\\nonumber \\
    \lc{R}_{0i0j}(Y_{\left|m\right|}) & =\left(G_{\text{Tele},I_{2}}-2G_{4,\phi}\right)\left(\begin{array}{ccc}
    \bigl(m^{2}+k{}^{2}\bigr)B_{1} & 0 & 0\\
    0 & \bigl(m^{2}+k{}^{2}\bigr)B_{1} & 0\\
    0 & 0 & m^{2}B_{1}
    \end{array}\right)\,.\label{eq:ERiemcase1massive}
\end{align}
The system in total assumes 7 PDoF which are divided as 6 in the massless sector described by Eq.~(\ref{eq:case1massless}) and 1 in the massive sector described by Eq.~(\ref{eq:case1massive}). The massless sector is parametrized by $\left(A_{1},..,A_{6}\right)$ which are packed as 2 scalars $\left(A_{1},A_{2}\right)$, one vector $\left(A_{3},A_{4}\right)$ and the tensor modes $\left(A_{5},A_{6}\right)$. Regarding the polarization content, the massless sector enjoys the usual tensor $(A_{5},A_{6})$ polarizations along with the breathing $\left(A_{1}\right)$ mode. The massive sector contains one massive scalar described by $B_{1}$. One can see that by setting $\left(G_{\text{Tele},I_{2}}-2G_{4,\phi}\right)\rightarrow0$
in Eq.~(\ref{eq:ERiemcase1massive}) the massive scalar mode becomes undetectable to the polarization detectors although it will still propagating as can be seen from Eq.~(\ref{eq:case1massive}).\\ \\
\textbf{\underline{Case 2}} $( G_{\text{Tele},T_{\text{vec}}}\neq0,\,G_{\text{Tele},T_{\text{ax}}}\neq0,\,\tilde{c}_{1}=0,\,\tilde{c}_{2}=0)$\\ \\
A pair of solutions of the system $ \tilde{c}_{1}=0,\tilde{c}_{2}=0 $
that covers the whole solution manifold are $$ G_{\text{Tele},T_{\text{vec}}}=-\tfrac{2}{3}(G_{4}-G_{\text{Tele},T}),G_{4,\phi}=\tfrac{1}{2}G_{\text{Tele},I_{2}} $$
and $$ G_{3,\phi}=\frac{\bigl(3(G_{\text{Tele},I_{2}}-2G_{4,\phi})^{2}+2(2(G_{4}-G_{\text{Tele},T})+3(G_{\text{Tele},T_{\text{vec}}}))(G_{\text{Tele},X}+G_{2,X})\bigr)}{4(2(G_{4}-G_{\text{Tele},T})+3(G_{\text{Tele},T_{\text{vec}}}))},\,G_{2,\phi\phi}=-G_{\text{Tele},\phi\phi} \,.$$\\ \\
\textbf{\underline{Case 2.I}}  $( G_{\text{Tele},T_{\text{vec}}}=-\tfrac{2}{3}(G_{4}-G_{\text{Tele},T}),\,G_{4,\phi}=\tfrac{1}{2}G_{\text{Tele},I_{2}}) $\\ \\
The determinant of the system reads as
\begin{align}
    P(k) & =2^{17}3^{-9}(G_{4}-G_{\text{Tele},T}){}^{8}(G_{\text{Tele},T_{\text{ax}}}){}^{3}k{}^{12}\bigl(\omega^{2}-k{}^{2}\bigr)^{7}\Bigl(\tilde{c}_{3}+\tilde{c}_{4}\bigl(\omega^{2}-k{}^{2}\bigr)\Bigr)\,,
\end{align}
which in turn means that we have more subcases.\\ \\
\textbf{\underline{Case 2.I.a}} $( G_{\text{Tele},T_{\text{ax}}}\neq0,\,\tilde{c}_{3}\neq0,\tilde{c}_{4}\neq0)$\\ \\
The corresponding important quantities in this subcase are:
\begin{align}
    P(k) & =\tfrac{131072}{19683}(G_{4}-G_{\text{Tele},T}){}^{8}G_{\text{Tele},T_{\text{ax}}}{}^{3}k{}^{12}\bigl(\omega^{2}-k{}^{2}\bigr)^{7}\Bigl(\tilde{c}_{3}+\tilde{c}_{4}\bigl(\omega^{2}-k{}^{2}\bigr)\Bigr)\,,\\\nonumber \\
    Y_{\omega} & =\left(\delta\phi,\Phi,\chi,\sigma,\beta{}_{i},\Sigma{}_{i},\Lambda{}_{i},h{}_{ij}\right)\,,\\\nonumber \\
    Y_{\left|k\right|} & =\left(0,0,0,A_{1},-A_{2},-A_{3},i\left|k\right|A_{2},i\left|k\right|A_{3},A_{2},A_{3},\frac{2A_{4}}{k^{2}},\frac{2A_{5}}{k^{2}}\right)^{T}\,,\\\nonumber \\
    \lc{R}_{0i0j}(Y_{\left|k\right|}) & =\left(\begin{array}{ccc}
    A_{4} & A_{5} & 0\\
    A_{5} & -A_{4} & 0\\
    0 & 0 & 0
    \end{array}\right)\,,\\\nonumber \\
    Y_{\left|m\right|} & =\left(B_{1},0,0,0,0,0,0,0,0,0,0,0\right)^{T}\,,\\\nonumber \\
    \lc{R}_{0i0j}(Y_{\left|m\right|}) & =\left(\begin{array}{ccc}
    0 & 0 & 0\\
    0 & 0 & 0\\
    0 & 0 & 0
    \end{array}\right)\,.
\end{align}
This is a peculiar case where the only scalar PDoF that are scalars $\sigma$ which is massless and $\delta\phi$ which is the massive are not at all involved in the polarizations. This is due to the fact that they are not coupled to any metric$\left(\Phi,\chi\right)$ scalar DoF and hence they cannot have a polarization imprint in both the massless and massive sectors.\\ \\
\textbf{\underline{Case 2.I.b}} $( G_{\text{Tele},T_{\text{ax}}}\neq0,\,\tilde{c}_{3}=0,\,\tilde{c}_{4}=0) $\\ \\
The solution of $ \tilde{c}_{3}=0,\tilde{c}_{4}=0 $
is $$ G_{3,\phi}=\frac{(G_{\text{Tele},X}+G_{2,X})}{2}\,,\quad G_{2,\phi\phi}=-G_{\text{Tele},\phi\phi} \,,$$ and then the important quantities become:
\begin{align}
    P(k) & =\tfrac{131072}{19683}(G_{4}-G_{\text{Tele},T}){}^{8}G_{\text{Tele},T_{\text{ax}}}{}^{3}k{}^{12}\bigl(\omega^{2}-k{}^{2}\bigr)^{7}\,,\\\nonumber \\
    Y_{\omega} & =\left(\Phi,\chi,\sigma,\beta{}_{i},\Sigma{}_{i},\Lambda{}_{i},h{}_{ij}\right)\,,\\\nonumber \\
    Y_{\left|k\right|} & =\left(0,0,A_{1},-A_{2},-A_{3},i\left|k\right|A_{2},i\left|k\right|A_{3},A_{2},A_{3},,\frac{2A_{4}}{k^{2}},\frac{2A_{5}}{k^{2}}\right)^{T}\,,\\\nonumber \\
    \lc{R}_{0i0j}(Y_{\left|k\right|}) & =\left(\begin{array}{ccc}
    A_{4} & A_{5} & 0\\
    A_{5} & -A_{4} & 0\\
    0 & 0 & 0
    \end{array}\right)\,.
\end{align}
\\ \\
\noindent \textbf{\underline{Case 2.II}} $( G_{3,\phi}=Z_2,\, G_{2,\phi\phi}=-G_{\text{Tele},\phi\phi}) $
\\ \\
In this case $Z_2$ is defined in~\eqref{Z2}. Further we have $2(G_{4}-G_{\text{Tele},T})+3G_{\text{Tele},T_{\text{vec}}}\neq0$ and then we have:
\begin{align}
    P(k) & =\tfrac{16384}{243}(G_{4}-G_{\text{Tele},T}){}^{5}G_{\text{Tele},T_{\text{vec}}}{}^{3}G_{\text{Tele},T_{\text{ax}}}{}^{3}(G_{\text{Tele},I_{2}}-2G_{4,\phi})k{}^{12}\bigl(\omega^{2}-k{}^{2}\bigr)^{8}\,.
\end{align}
There are further subsubcases:\\ \\
\textbf{\underline{Case 2.II.a}} ($G_{\text{Tele},I_{2}}-2G_{4,\phi}\neq0.$)\\ \\
Here, we have the following terms:
\begin{align}
    P(k) & =\tfrac{16384}{243}(G_{4}-G_{\text{Tele},T}){}^{5}(G_{\text{Tele},T_{\text{vec}}}){}^{3}(G_{\text{Tele},T_{\text{ax}}}){}^{3}(G_{\text{Tele},I_{2}}-2G_{4,\phi})k{}^{12}\bigl(\omega^{2}-k{}^{2}\bigr)^{8}\,,\\\nonumber \\
    Y_{\omega} & =\left(\delta\phi,\psi,\chi,\sigma,\beta{}_{i},\Sigma{}_{i},\Lambda{}_{i},h{}_{ij}\right)\,,\\\nonumber \\
    Y_{\left|k\right|} & =\left(-\frac{4(G_{4}-G_{\text{Tele},T}+G_{\text{Tele},T_{\text{vec}}})}{(G_{\text{Tele},I_{2}}-2G_{4,\phi})k{}^{2}}A_{1},-\frac{A_{1}}{k^{2}},\frac{iA_{1}}{k^{3}},A_{2},-A_{3},-A_{4},ikA_{3},ikA_{4},A_{3},A_{4},\frac{2A_{5}}{k^{2}},\frac{2A_{6}}{k^{2}}\right)^{T}\,,\\\nonumber \\
    \lc{R}_{0i0j}(Y_{\left|k\right|}) & =\left(\begin{array}{ccc}
    A_{1}+A_{5} & A_{6} & 0\\
    A_{6} & A_{1}-A_{5} & 0\\
    0 & 0 & 0
    \end{array}\right)\,.
\end{align}\\ \\
\textbf{\underline{Case 2.II.b}} ($G_{\text{Tele},I_{2}}-2G_{4,\phi}=0.$) \\ \\
The determinant of the system in this case is
\begin{align}
    P(k) & =-\tfrac{16384}{243}(G_{4}-G_{\text{Tele},T}){}^{4}G_{\text{Tele},T_{\text{vec}}}{}^{3}(G_{4}-G_{\text{Tele},T}+G_{\text{Tele},T_{\text{vec}}})G_{\text{Tele},T_{\text{ax}}}{}^{3}k{}^{12}\bigl(\omega^{2}-k{}^{2}\bigr)^{7}\,.
\end{align}\\
\textbf{\underline{Case 2.II.b.1}} $(G_{4}-G_{\text{Tele},T}+G_{\text{Tele},T_{\text{vec}}}\neq0)$ \\
\begin{align}
    P(k) & =-\tfrac{16384}{243}(G_{4}-G_{\text{Tele},T}){}^{4}G_{\text{Tele},T_{\text{vec}}}{}^{3}(G_{4}-G_{\text{Tele},T}+G_{\text{Tele},T_{\text{vec}}})G_{\text{Tele},T_{\text{ax}}}{}^{3}k{}^{12}\bigl(\omega^{2}-k{}^{2}\bigr)^{7}\,,\\\nonumber \\
    Y_{\omega} & =\left(\Phi,\chi,\sigma,\beta{}_{i},\Sigma{}_{i},\Lambda{}_{i},h{}_{ij}\right)\,,\\\nonumber \\
    Y_{\left|k\right|} & =\left(0,0,A_{1},-A_{2},-A_{3},i\left|k\right|A_{2},i\left|k\right|A_{3},A_{2},A_{3},\frac{2A_{4}}{k^{2}},\frac{2A_{5}}{k^{2}}\right)^{T}\,,\\\nonumber \\
    \lc{R}_{0i0j}(Y_{\left|k\right|}) & =\left(\begin{array}{ccc}
    A_{4} & A_{5} & 0\\
    A_{5} & -A_{4} & 0\\
    0 & 0 & 0
    \end{array}\right)\,.
\end{align}
\textbf{\underline{Case 2.II.b.2}} $(G_{4}-G_{\text{Tele},T}+G_{\text{Tele},T_{\text{vec}}}=0)$\\
\begin{align}
    P(k) & =-\tfrac{4096}{243}i(G_{4}-G_{\text{Tele},T}){}^{7}(G_{\text{Tele},T_{\text{ax}}}){}^{3}\omega k{}^{8}\bigl(\omega^{2}-k{}^{2}\bigr)^{7}\,,\\\nonumber \\
    Y_{\omega} & =\left(\Phi,\sigma,\beta{}_{i},\Sigma{}_{i},\Lambda{}_{i},h{}_{ij}\right)\,,\\\nonumber \\
    Y_{\left|k\right|} & =\left(0,A_{1},-A_{2},-A_{3},i\left|k\right|A_{2},i\left|k\right|A_{3},A_{2},A_{3},\frac{2A_{4}}{k^{2}},\frac{2A_{5}}{k^{2}}\right)^{T}\,,\\\nonumber \\
    \lc{R}_{0i0j}(Y_{\left|k\right|}) & =\left(\begin{array}{ccc}
    A_{4} & A_{5} & 0\\
    A_{5} & -A_{4} & 0\\
    0 & 0 & 0
    \end{array}\right)\,.
\end{align}
\\ \\
\textbf{\underline{Case 3}} $( G_{\text{Tele},T_{\text{vec}}}\neq0,\,G_{\text{Tele},T_{\text{ax}}}=0,\,\tilde{c}_{1}\neq0,\,\tilde{c}_{2}\neq0) $\\
\begin{align}
    P(k) & =-32i(G_{4}-G_{\text{Tele},T}){}^{5}G_{\text{Tele},T_{\text{vec}}}{}^{3}\omega^{2}k{}^{10}\bigl(\omega^{2}-k{}^{2}\bigr)^{3}\Bigl(\tilde{c}_{1}+\tilde{c}_{2}\bigl(\omega^{2}-k{}^{2}\bigr)\Bigr)\,,\\\nonumber \\
    Y_{\omega} & =\left(\delta\phi,\psi,\Phi,\chi,\beta{}_{i},\Sigma{}_{i},h{}_{ij}\right)\,,\\\nonumber \\
    Y_{\left|k\right|} & =\Big(0,-\frac{A_{1}}{k^{2}},-\frac{2(G_{4}-G_{\text{Tele},T}+G_{\text{Tele},T_{\text{vec}}})}{(G_{\text{Tele},T_{\text{vec}}})k{}^{2}}A_{1},\frac{i(2(G_{4}-G_{\text{Tele},T})+3G_{\text{Tele},T_{\text{vec}}})}{(G_{\text{Tele},T_{\text{vec}}})\left|k\right|{}^{3}}A_{1},0\nonumber\\
    &\,\,\,\,\,\,\,\,\,\,\,\,\,,0,0,0,\frac{2A_{2}}{k^{2}},\frac{2A_{3}}{k^{2}}\Big)^{T}\,,\\ \nonumber \\
    \lc{R}_{0i0j}(Y_{\left|k\right|}) & =\left(\begin{array}{ccc}
    A_{1}+A_{2} & A_{3} & 0\\
    A_{3} & A_{1}-A_{2} & 0\\
    0 & 0 & 0
    \end{array}\right)\,,\\\nonumber \\
    Y_{\left|m\right|} & =\left(-2(2(G_{4}-G_{\text{Tele},T})+3G_{\text{Tele},T_{\text{vec}}})B_{1},-(G_{\text{Tele},I_{2}}-2G_{4,\phi})B_{1},(G_{\text{Tele},I_{2}}-2G_{4,\phi})B_{1},0,0,0,0,0,0,0\right)^{T}\,,\\\nonumber \\
    \lc{R}_{0i0j}(Y_{\left|m\right|}) & =(G_{\text{Tele},I_{2}}-2G_{4,\phi})\left(\begin{array}{ccc}
    \bigl(m^{2}+k{}^{2}\bigr)B_{1} & 0 & 0\\
    0 & \bigl(m^{2}+k{}^{2}\bigr)B_{1} & 0\\
    0 & 0 & m^{2}B_{1}
    \end{array}\right)\,,
\end{align}
where we can hide the massive scalar from the polarizations by setting
$(G_{\text{Tele},I_{2}}-2G_{4,\phi})=0$.\\ \\
\textbf{\underline{Case 4}} $( G_{\text{Tele},T_{\text{vec}}}\neq0,\,G_{\text{Tele},T_{\text{ax}}}=0,\,\tilde{c}_{1}=0,\,\tilde{c}_{2}=0)$ \\ \\
A pair of solutions of the system $ \tilde{c}_{1}=0,\tilde{c}_{2}=0$
that cover the whole solution manifold are $$ G_{\text{Tele},T_{\text{vec}}}=-\tfrac{2}{3}(G_{4}-G_{\text{Tele},T})\,,\quad G_{4,\phi}=\tfrac{1}{2}G_{\text{Tele},I_{2}}\,,$$
and
$$ G_{3,\phi}=\frac{\bigl(3(G_{\text{Tele},I_{2}}-2G_{4,\phi})^{2}+2(2(G_{4}-G_{\text{Tele},T})+3G_{\text{Tele},T_{\text{vec}}})(G_{\text{Tele},X}+G_{2,X})\bigr)}{4(2(G_{4}-G_{\text{Tele},T})+3G_{\text{Tele},T_{\text{vec}}})}\,,\quad G_{2,\phi\phi}=-G_{\text{Tele},\phi\phi}\,.$$\\ \\
\textbf{\underline{Case 4.I}} $( G_{\text{Tele},T_{\text{vec}}}=-\tfrac{2}{3}(G_{4}-G_{\text{Tele},T}),\,G_{4,\phi}=\tfrac{1}{2}G_{\text{Tele},I_{2}}) $\\ \\
By defining
\begin{align}
    \tilde{c}_{3} & =-G_{\text{Tele},\phi\phi}-G_{2,\phi\phi}\,,\\
    \tilde{c}_{4} & =G_{\text{Tele},X}+G_{2,X}-2G_{3,\phi}\,,
\end{align}
the determinant of the system reads as
\begin{align}
    P(k) & =\tfrac{256}{81}i(G_{4}-G_{\text{Tele},T}){}^{8}\omega^{2}k^{10}\bigl(\omega^{2}-k^{2}\bigr)^{2}\Bigl(\tilde{c}_{3}+\tilde{c}_{4}\bigl(\omega^{2}-k^{2}\bigr)\Bigr)\,,
\end{align}
which in turn means that we have more subcases\\ \\

\textbf{\underline{Case 4.I.a}} $(\tilde{c}_{3}\neq0,\,\tilde{c}_{4}\neq0)$\\
\begin{align}
    P(k) & =\tfrac{256}{81}i(G_{4}-G_{\text{Tele},T}){}^{8}\omega^{2}k{}^{10}\bigl(\omega^{2}-k{}^{2}\bigr)^{2}\Bigl(\tilde{c}_{3}+\tilde{c}_{4}\bigl(\omega^{2}-k{}^{2}\bigr)\Bigr)\,,\\\nonumber \\
    Y_{\omega} & =\left(\delta\phi,\psi,\chi,\beta{}_{i},\Sigma{}_{i},h{}_{ij}\right)\,,\\\nonumber \\
    Y_{\left|k\right|} & =\left(0,0,0,0,0,0,0,\frac{2}{k^{2}}A_{1},\frac{2}{k^{2}}A_{2}\right)^{T}\,,\\\nonumber \\
    \lc{R}_{0i0j}(Y_{\left|k\right|}) & =\left(\begin{array}{ccc}
    A_{1} & A_{2} & 0\\
    A_{2} & -A_{1} & 0\\
    0 & 0 & 0
    \end{array}\right)\,,\\\nonumber \\
    Y_{\left|m\right|} & =\left(B_{1},0,0,0,0,0,0,0,0\right)^{T}\,,\\\nonumber \\
    \lc{R}_{0i0j}(Y_{\left|m\right|}) & =\left(\begin{array}{ccc}
    0 & 0 & 0\\
    0 & 0 & 0\\
    0 & 0 & 0
    \end{array}\right)\,.
\end{align}\\
 \textbf{\underline{Case 4.I.b}} $( \tilde{c}_{3}=0,\,\tilde{c}_{4}=0)$\\ \\
The solution of $\tilde{c}_{3}=0,\,\tilde{c}_{4}=0 $
is $$G_{3,\phi}=\tfrac{1}{2}(G_{\text{Tele},X}+G_{2,X})\,,\quad G_{2,\phi\phi}=-G_{\text{Tele},\phi\phi} \,,$$
and then we have:
\begin{align}
    P(k) & =\tfrac{256}{81}i(G_{4}-G_{\text{Tele},T}){}^{8}\omega^{2}k{}^{10}\bigl(\omega^{2}-k{}^{2}\bigr)^{2}\,,\\\nonumber \\
    Y_{\omega} & =\left(\psi,\chi,\beta{}_{i},\Sigma{}_{i},h{}_{ij}\right)\,,\\\nonumber \\
    Y_{\left|k\right|} & =\left(0,0,0,0,0,0,\frac{2}{k^{2}}A_{1},\frac{2}{k^{2}}A_{2}\right)^{T}\,,\\\nonumber \\
    \lc{R}_{0i0j}(Y_{\left|k\right|}) & =\left(\begin{array}{ccc}
    A_{1} & A_{2} & 0\\
    A_{2} & -A_{1} & 0\\
    0 & 0 & 0
    \end{array}\right)\,.
\end{align}
\\ \\
\textbf{\underline{Case 4.II}} $(G_{3,\phi}=Z_2,\, G_{2,\phi\phi}=-G_{\text{Tele},\phi\phi},\, 2(G_{4}-G_{\text{Tele},T})+3G_{\text{Tele},T_{\text{vec}}}\neq0) $\\ \\
In this case with $Z_2$ defined in~\eqref{Z2} we have
\begin{align}
    P(k) & =-32i(G_{4}-G_{\text{Tele},T}){}^{5}(G_{\text{Tele},T_{\text{vec}}}){}^{3}(G_{\text{Tele},I_{2}}-2G_{4,\phi})\omega^{2}k{}^{10}\bigl(\omega^{2}-k{}^{2}\bigr)^{3}\,,
\end{align}
and then more subcases appear.\\ \\
\textbf{\underline{Case 4.II.a}} $(G_{\text{Tele},I_{2}}-2G_{4,\phi}\neq0)$\\
\begin{align}
    P(k) & =-32i(G_{4}-G_{\text{Tele},T}){}^{5}G_{\text{Tele},T_{\text{vec}}}{}^{3}(G_{\text{Tele},I_{2}}-2G_{4,\phi})\omega^{2}k{}^{10}\bigl(\omega^{2}-k{}^{2}\bigr)^{3}\,,\\\nonumber \\
    Y_{\omega} & =\left(\delta\phi,\psi,\chi,\beta{}_{i},\Sigma{}_{i},h{}_{ij}\right)\,,\\\nonumber \\
    Y_{\left|k\right|} & =\left(-\frac{4(G_{4}-G_{\text{Tele},T}+G_{\text{Tele},T_{\text{vec}}})}{(G_{\text{Tele},I_{2}}-2G_{4,\phi})k{}^{2}}A_{1},-\frac{A_{1}}{k^{2}},\frac{iA_{1}}{\left|k\right|{}^{3}},0,0,0,0,\frac{2A_{2}}{k^{2}},\frac{2A_{3}}{k^{2}}\right)^{T}\,,\\\nonumber \\
    \lc{R}_{0i0j}(Y_{\left|k\right|}) & =\left(\begin{array}{ccc}
    A_{1}+A_{2} & A_{3} & 0\\
    A_{3} & A_{1}-A_{2} & 0\\
    0 & 0 & 0
    \end{array}\right)\,.
\end{align}\\
\textbf{\underline{Case 4.II.b}} $(G_{\text{Tele},I_{2}}-2G_{4,\phi}=0)$\\ \\
The determinant of the system in this case is
\begin{align}
    P(k) & =32i(G_{4}-G_{\text{Tele},T}){}^{4}G_{\text{Tele},T_{\text{vec}}}{}^{3}(G_{4}-G_{\text{Tele},T}+G_{\text{Tele},T_{\text{vec}}})\omega^{2}k{}^{10}\bigl(\omega^{2}-k{}^{2}\bigr)^{2}\,.
\end{align}
\\
\textbf{\underline{Case 4.II.b.1}} $(G_{4}-G_{\text{Tele},T}+G_{\text{Tele},T_{\text{vec}}}\neq0)$\\
\begin{align}
    P(k) & =32i(G_{4}-G_{\text{Tele},T}){}^{4}G_{\text{Tele},T_{\text{vec}}}{}^{3}(G_{4}-G_{\text{Tele},T}+G_{\text{Tele},T_{\text{vec}}})\omega^{2}k{}^{10}\bigl(\omega^{2}-k{}^{2}\bigr)^{2}\,,\\\nonumber \\
    Y_{\omega} & =\left(\psi,\chi,\beta{}_{i},\Sigma{}_{i},h{}_{ij}\right)\,,\\\nonumber \\
    Y_{\left|k\right|} & =\left(0,0,0,0,0,0,\frac{2A_{1}}{k^{2}},\frac{2A_{2}}{k^{2}}\right)^{T}\,,\\\nonumber \\
    \lc{R}_{0i0j}(Y_{\left|k\right|}) & =\left(\begin{array}{ccc}
    A_{1} & A_{2} & 0\\
    A_{2} & -A_{1} & 0\\
    0 & 0 & 0
    \end{array}\right)\,.
\end{align}
 \textbf{\underline{Case 4.II.b.2}} $(G_{4}-G_{\text{Tele},T}+G_{\text{Tele},T_{\text{vec}}}=0)$\\
\begin{align}
    P(k) & =-8(G_{4}-G_{\text{Tele},T}){}^{7}\omega^{3}k{}^{6}\bigl(\omega^{2}-k{}^{2}\bigr)^{2}\,,\\\nonumber \\
    Y_{\omega} & =\left(\psi,\beta{}_{i},\Sigma{}_{i},h{}_{ij}\right)\,,\\\nonumber \\
    Y_{\left|k\right|} & =\left(0,0,0,0,0,\frac{2A_{1}}{k^{2}},\frac{2A_{2}}{k^{2}}\right)^{T}\,,\\\nonumber \\
    \lc{R}_{0i0j}(Y_{\left|k\right|}) & =\left(\begin{array}{ccc}
    A_{1} & A_{2} & 0\\
    A_{2} & -A_{1} & 0\\
    0 & 0 & 0
    \end{array}\right)\,.
\end{align}
\\ \\
\textbf{\underline{Case 5}} $( G_{\text{Tele},T_{\text{vec}}}=0,\,G_{\text{Tele},T_{\text{ax}}}\neq0,\,\tilde{c}_{1}\neq0,\,\tilde{c}_{2}\neq0)$\\ \\
In this case we find the following quantities:
\begin{align}
    P(k) & =-\tfrac{2048}{243}i(G_{4}-G_{\text{Tele},T}){}^{5}G_{\text{Tele},T_{\text{ax}}}{}^{3}\omega^{2}k{}^{10}\bigl(\omega^{2}-k{}^{2}\bigr)^{3}\Bigl(\tilde{c}_{1}+\tilde{c}_{2}\bigl(\omega^{2}-k{}^{2}\bigr)\Bigr)\,,\\\nonumber \\
    Y_{\omega} & =\left(\delta\phi,\psi,\Phi,\sigma,\beta{}_{i},\Sigma{}_{i},h{}_{ij}\right)\,,\\\nonumber \\
    Y_{\left|k\right|} & =\left(0,0,0,A_{1},0,0,0,0,\frac{2A_{2}}{k^{2}},\frac{2A_{3}}{k^{2}}\right)^{T}\,,\\\nonumber \\
    \lc{R}_{0i0j}(Y_{\left|k\right|}) & =\left(\begin{array}{ccc}
    A_{2} & A_{3} & 0\\
    A_{3} & -A_{2} & 0\\
    0 & 0 & 0
    \end{array}\right)\,,\\\nonumber \\
    Y_{\left|m\right|} & =\left(-4(G_{4}-G_{\text{Tele},T})B_{1},-(G_{\text{Tele},I_{2}}-2G_{4,\phi})B_{1},(G_{\text{Tele},I_{2}}-2G_{4,\phi})B_{1},0,0,0,0,0,0,0\right)^{T}\,,\\\nonumber \\
    \lc{R}_{0i0j}(Y_{\left|m\right|}) & =\left(G_{\text{Tele},I_{2}}-2G_{4,\phi}\right)\left(\begin{array}{ccc}
    \bigl(m^{2}+k{}^{2}\bigr)B_{1} & 0 & 0\\
    0 & \bigl(m^{2}+k{}^{2}\bigr)B_{1} & 0\\
    0 & 0 & m^{2}B_{1}
    \end{array}\right)\,.
\end{align}
\\ \\
\textbf{\underline{Case 6:}} $(G_{\text{Tele},T_{\text{vec}}}=0,\,G_{\text{Tele},T_{\text{ax}}}\neq0,\,\tilde{c}_{1}=0,\,\tilde{c}_{2}=0)$\\ \\
A solution set of the system that covers the whole solution manifold is
\begin{equation}
 G_{3,\phi}=\frac{3(G_{\text{Tele},I_{2}}-2G_{4,\phi})^{2}}{8(G_{4}-G_{\text{Tele},T})}+\tfrac{1}{2}(G_{\text{Tele},X}+G_{2,X}),G_{2,\phi\phi}=-G_{\text{Tele},\phi\phi}\,,
\end{equation}
which leads us to the determinant
\begin{align}
    P(k) & =-\tfrac{2048}{243}i(G_{4}-G_{\text{Tele},T}){}^{5}G_{\text{Tele},T_{\text{ax}}}{}^{3}(G_{\text{Tele},I_{2}}-2G_{4,\phi})\omega^{2}k{}^{10}\bigl(\omega^{2}-k{}^{2}\bigr)^{3}\,,
\end{align}
which leads several subcases:\\ \\
\textbf{\underline{Case 6.I}} $(G_{\text{Tele},I_{2}}-2G_{4,\phi}\neq0)$\\
\begin{align}
    P(k) & =-\tfrac{2048}{243}i(G_{4}-G_{\text{Tele},T}){}^{5}G_{\text{Tele},T_{\text{ax}}}{}^{3}(G_{\text{Tele},I_{2}}-2G_{4,\phi})\omega^{2}k{}^{10}\bigl(\omega^{2}-k{}^{2}\bigr)^{3}\,,\\\nonumber \\
    Y_{\omega} & =\left(\delta\phi,\psi,\sigma,\beta{}_{i},\Sigma{}_{i},h{}_{ij}\right)\,,\\\nonumber \\
    Y_{\left|k\right|} & =\left(0,0,A_{1},0,0,0,0,\frac{2A_{2}}{k^{2}},\frac{2A_{3}}{k^{2}}\right)^{T}\,,\\\nonumber \\
    \lc{R}_{0i0j}(Y_{\left|k\right|}) & =\left(\begin{array}{ccc}
    A_{2} & A_{3} & 0\\
    A_{3} & -A_{2} & 0\\
    0 & 0 & 0
    \end{array}\right)\,.
\end{align}
\textbf{\underline{Case 6.II}} $(G_{\text{Tele},I_{2}}-2G_{4,\phi}=0)$\\
\begin{align}
    P(k) & =\tfrac{2048}{243}i(G_{4}-G_{\text{Tele},T}){}^{5}G_{\text{Tele},T_{\text{ax}}}{}^{3}\omega^{2}k{}^{8}\bigl(\omega^{2}-k{}^{2}\bigr)^{3}\,,\\\nonumber \\
    Y_{\omega} & =\left(\psi,\sigma,\beta{}_{i},\Sigma{}_{i},h{}_{ij}\right)\,,\\\nonumber \\
    Y_{\left|k\right|} & =\left(0,A_{1},0,0,0,0,\frac{2A_{2}}{k^{2}},\frac{2A_{3}}{k^{2}}\right)^{T}\,,\\\nonumber \\
    \lc{R}_{0i0j}(Y_{\left|k\right|}) & =\left(\begin{array}{ccc}
    A_{2} & A_{3} & 0\\
    A_{3} & -A_{2} & 0\\
    0 & 0 & 0
    \end{array}\right)\,.
\end{align}
\\ \\
\textbf{\underline{Case 7:}}  $( G_{\text{Tele},T_{\text{vec}}}=0,\,G_{\text{Tele},T_{\text{ax}}}=0,\,\tilde{c}_{1}\neq0,\,\tilde{c}_{2}\neq0)$\\ \\
In this case we obtain:
\begin{align}
    P(k) & =-4(G_{4}-G_{\text{Tele},T}){}^{5}(\omega^{2}-k{}^{2})^{2}k{}^{8}\Bigl(\tilde{c}_{1}+\tilde{c}_{2}\bigl(\omega^{2}-k{}^{2}\bigr)\Bigr)\,,\\\nonumber \\
    Y_{\omega} & =\left(\delta\phi,\psi,\Phi,\beta{}_{i},h{}_{ij}\right)\,,\\\nonumber \\
    Y_{\left|k\right|} & =\left(0,0,0,0,0,\frac{2A_{1}}{k^{2}},\frac{2A_{2}}{k^{2}}\right)^{T}\,,\\\nonumber \\
    \lc{R}_{0i0j}(Y_{\left|k\right|}) & =\left(\begin{array}{ccc}
    A_{1} & A_{2} & 0\\
    A_{2} & -A_{1} & 0\\
    0 & 0 & 0
    \end{array}\right)\,,\\\nonumber \\
    Y_{\left|m\right|} & =\left(-4(G_{4}-G_{\text{Tele},T})B_{1},-(G_{\text{Tele},I_{2}}-2G_{4,\phi})B_{1},(G_{\text{Tele},I_{2}}-2G_{4,\phi})B_{1},0,0,0,0\right)^{T}\,,\\\nonumber \\
    \lc{R}_{0i0j}(Y_{\left|m\right|}) & =(G_{\text{Tele},I_{2}}-2G_{4,\phi})\left(\begin{array}{ccc}
    \bigl(m^{2}+k{}^{2}\bigr)B_{1} & 0 & 0\\
    0 & \bigl(m^{2}+k{}^{2}\bigr)B_{1} & 0\\
    0 & 0 & m^{2}B_{1}
    \end{array}\right)\,.
\end{align}
This case gives us exactly the full Horndeski Case 0i. in the limit $G_{\text{Tele},I_{2}}\rightarrow0$. Thus the full Horndeski theory is a sub-branch of Case 7 and just like we discussed in~\ref{Case-0ii.}  with $\left(G_{\text{Tele},I_{2}}-2G_{4,\phi}\right)\rightarrow0$ where it
hides the massive scalar PDoF from the polarization detectors.
\\ \\
\textbf{\underline{Case 8:}} $( G_{\text{Tele},T_{\text{vec}}}=0,\, G_{\text{Tele},T_{\text{ax}}}=0,\,\tilde{c}_{1}=0,\,\tilde{c}_{2}=0) $\\ \\
A solution set of this system that covers the whole solution manifold is
\begin{equation}
     G_{3,\phi}=\frac{3(G_{\text{Tele},I_{2}}-2G_{4,\phi})^{2}}{8(G_{4}-G_{\text{Tele},T})}+\tfrac{1}{2}(G_{\text{Tele},X}+G_{2,X})\,,\quad G_{2,\phi\phi}=-G_{\text{Tele},\phi\phi}\,,
\end{equation}
which leads us to the determinant
\begin{align}
    P(k) & =4(G_{4}-G_{\text{Tele},T}){}^{5}(G_{\text{Tele},I_{2}}-2G_{4,\phi})k{}^{8}\bigl(\omega^{2}-k{}^{2}\bigr)^{2}\,.
\end{align}
The Horndeski Case 0ii. is a subcase for $G_{\text{Tele},I_{2}}\rightarrow0$.\\ \\
\textbf{\underline{Case 8.I}} $(G_{\text{Tele},I_{2}}-2G_{4,\phi}\neq0)$\\
\begin{align}
    P(k) & =4(G_{4}-G_{\text{Tele},T}){}^{5}(G_{\text{Tele},I_{2}}-2G_{4,\phi})k{}^{8}\bigl(\omega^{2}-k{}^{2}\bigr)^{2}\,,\\\nonumber \\
    Y_{\omega} & =\left(\delta\phi,\psi,\beta{}_{i},h{}_{ij}\right)\,,\\\nonumber \\
    Y_{\left|k\right|} & =\left(0,0,0,0,\frac{2A_{1}}{k^{2}},\frac{2A_{2}}{k^{2}}\right)^{T}\,,\\\nonumber \\
    \lc{R}_{0i0j}(Y_{\left|k\right|}) & =\left(\begin{array}{ccc}
    A_{1} & A_{2} & 0\\
    A_{2} & -A_{1} & 0\\
    0 & 0 & 0
    \end{array}\right)\,.
\end{align}
\textbf{\underline{Case 8.II}} $(G_{\text{Tele},I_{2}}-2G_{4,\phi}=0)$\\
\begin{align}
    P(k) & =-4(G_{4}-G_{\text{Tele},T}){}^{5}k{}^{6}\bigl(\omega^{2}-k{}^{2}\bigr)^{2}\,,\\\nonumber \\
    Y_{\omega} & =\left(\psi,\beta{}_{i},h{}_{ij}\right)\,,\\\nonumber \\
    Y_{\left|k\right|} & =\left(0,0,0,\frac{2A_{1}}{k^{2}},\frac{2A_{2}}{k^{2}}\right)^{T}\,,\\\nonumber \\
    \lc{R}_{0i0j}(Y_{\left|k\right|}) & =\left(\begin{array}{ccc}
    A_{1} & A_{2} & 0\\
    A_{2} & -A_{1} & 0\\
    0 & 0 & 0
    \end{array}\right)\,.
\end{align}

\bibliographystyle{utphys}
\bibliography{references}

\end{document}